\documentclass[review,fleqn,authoryear]{elsarticle}
\usepackage{amssymb,amsbsy,amsthm,amsmath,amsfonts,amssymb,amscd}
\usepackage{geometry}
\usepackage{babel,blindtext}
\usepackage{lipsum}
\usepackage{bm} % for \bm macro
\usepackage{graphicx}
\usepackage{tabularx}
\graphicspath{{./images/}}
\usepackage{float}
\usepackage{booktabs, multirow, makecell}
\usepackage{caption}
\usepackage{subcaption}
\usepackage{multirow}
\usepackage{cancel}
\usepackage{lineno}
\usepackage{hyperref}
\usepackage{graphicx}
\usepackage{mathrsfs}
\usepackage{booktabs}
\hypersetup{colorlinks, citecolor=black, filecolor=black, linkcolor=black, urlcolor=black}
\usepackage{booktabs}
\usepackage[ruled,linesnumbered]{algorithm2e}
\usepackage[toc,page,titletoc]{appendix}
\usepackage[capitalise,nameinlink]{cleveref}
\crefname{supp}{Supplement}{Supplements}
\usepackage{algpseudocode}
\usepackage{mathtools,nccmath}
\usepackage{adjustbox}
\usepackage{cleveref}
\usepackage{xparse}
\theoremstyle{definition}

\usepackage{color}
\usepackage{xcolor}
\usepackage{amsmath}
\usepackage{lineno}
\usepackage{soul}
\colorlet{soulred}{red!40}

\definecolor{airforceblue}{rgb}{0.36, 0.54, 0.66}

\makeatletter
\renewcommand{\paragraph}{\@startsection{paragraph}{4}{0pt}%
  {3.25ex plus 1ex minus 0.2ex}{0.75ex}%
  {\normalsize\itshape}}
\makeatother

\begin{document}
%%%%%%%%%%%%%%%%%%%%%%%%%%%%%%%%%%%%%%%%%%%%%%%%%%%

\begin{frontmatter}
\title{BRIDGE: Bridging Routine Inputs and Digital modelling for Geotechnical Engineering}

\author{Roshan Philip Saji ${}^{1}$ \corref{cor1}}
\author{Waleed El-Sekelly ${}^{1,2}$}
\author{Tarek Abdoun ${}^{1}$}
\author{Mostafa E. Mobasher ${}^{1}$ \corref{cor2}}

\cortext[cor1]{Corresponding author. \emph{E-mail address:} \texttt{rs7625@nyu.edu} (Roshan Philip Saji)}

\cortext[cor2]{Corresponding author. \emph{E-mail address:} \texttt{mostafa.mobasher@nyu.edu} (Mostafa E. Mobasher)}

\address{${}^{1}$ Civil and Urban Engineering Department, New York University Abu Dhabi, Abu Dhabi, P.O. Box 129188, UAE}
\address{${}^{2}$Department of Structural Engineering, Mansoura University, Mansoura, P.O. Box 35516, Egypt}

\begin{abstract}
The mechanical properties needed for geotechnical design require laboratory
tests and are thus available for far fewer field samples than routinely
measured basic (index) properties which are available for nearly every sample. We term this difference in availability between the two classes of properties as the characterization gap. Purely data-driven models do not explicitly enforce physical laws and may be unreliable beyond their training domain, while direct constitutive calibration preserves physical consistency but requires advanced laboratory data and expert judgment. We therefore propose \textbf{BRIDGE} (\textbf{B}ridging \textbf{R}outine \textbf{I}nputs and \textbf{D}igital modelling for \textbf{G}eotechnical \textbf{E}ngineering) a modular framework that predicts mechanical properties from routinely available basic properties. BRIDGE screens the dataset for outliers and non-physical responses, bins the retained samples into Cells of comparable properties, and selects a small calibration dataset from each Cell, while the remaining samples are used for validation. A mechanical model is then
calibrated for each calibration sample, and the converged calibrations collectively define the mechanical surrogate of the Cell together with its parameter ranges and the corresponding predictive envelope in the mechanical property space. A Cell is validated when most validation samples fall within this envelope, whereupon the surrogate can predict the mechanical property ranges of new samples whose basic properties lie within the Cell. We adopt a finite element Extended Cam-Clay (ECC) model, whose four parameters are calibrated in a multi-stage process. Applied to the marine clay database Tokyo-CLAY/14/67760, BRIDGE predicts the mechanical properties of an entire Cell from calibrations of just 10 to 20\% of its samples. Finally, since the mechanical model, the binning criteria, and the sampling method are incorporated into BRIDGE as independent and interchangeable components, the framework is not restricted to the present configuration and is readily extensible to other constitutive models, geomaterial databases, and engineering applications.
\end{abstract}

\end{frontmatter}

\section{Introduction}
\label{Sec:Introduction}

\subsection{Motivation}
\label{sec:motivation}

%%%%%%%%%%%%%%%%%%%%%%%%%%%%%%%%%%%%%%
Geotechnical design requires mechanical properties describing soil strength, stiffness, and compressibility, including the undrained shear strength $S_u$, undrained secant modulus $E_{50}$, preconsolidation pressure $p_o$, and compression index $C_c$ \citep{ladd1974, jamiolkowski1985new, kulhawy1990manual, lunne1997cone, mayne2002subsurface}. These properties are commonly obtained through time and resource intensive experiments like triaxial and oedometer tests and are therefore measured for relatively few samples \citep{astm_d4767, astm_d2435, schumacher2023challenges}. By contrast, the basic properties, also commonly referred to as index properties, such as the liquid limit $LL$, plastic limit $PL$, water content $w$, and void ratio $e$ are routinely available for much larger sample sets \citep{astm_d4318, astm_d2216, holtz2011introduction}. We define this disparity between the limited availability of mechanical properties required for design and the broader availability of basic properties as the \textit{characterization gap}. Among the mechanical properties considered here, $S_u$ is more commonly available because it can be estimated rapidly in the field using a pocket penetrometer or vane shear tests \citep{astm_d8121, mayne2002subsurface}, thus
making it a practical classification tool for soils \citep{terzaghi1996soil, iso14688_2}. Ideally, a method that addresses the characterization gap should 
\begin{enumerate}[(\alph*)]
    \item predict the mechanical properties from the routinely available
          basic properties,
    \item maintain physical consistency within and beyond the
          data from which it is derived, and
    \item require only limited experimental or training data.
\end{enumerate}

%%%%%%%%%%%%%%%%%%%%%%%%%%%%%%%%%%%%%%
\subsection{Existing Approaches and Their Limitations}
\label{sec:approaches}

We organize the existing literature that address the characterization gap into three categories, namely data-driven approaches, direct physics-based approaches  and hybrid approaches. Each category is reviewed below and assessed against requirements (a), (b), and (c) mentioned above.

\subsubsection{Data-Driven Approaches}
\label{sec:datadriven}

Data-driven approaches predict the mechanical properties directly from the basic properties without solving a mechanical model, and comprise of empirical correlations and machine learning models. Classical empirical correlations relate the Atterberg limits and other routine measurements to quantities such as the compression index and the normalized undrained shear strength \citep{skempton1944notes, skempton1957discussion, wroth1978correlation, koutsoftas1985design}, and their simplicity and interpretability make them useful for preliminary geomaterial characterization. However, the underlying relationships are influenced by mineralogy, stress history, fabric, anisotropy, and sampling quality \citep{dolinar2010predicting, shimobe2021relationships}. For example, the undrained strength at the plastic limit alone varies by more than an order of magnitude across fine-grained soils due to differences in clay mineralogy \citep{haigh2013plastic, lehmann2021clays}. Empirical correlations therefore address requirement~(a), but they are not reliably consistent beyond their source data, thus not satisfying requirement~(b).

Machine-learning models have also been used to close the characterization gap by capturing the nonlinear, multivariate relationships between basic properties and mechanical properties, particularly $S_u$. Models using the Atterberg limits, water content, unit weight, and stress history have been reported to outperform classical correlations \citep{zhang2021prediction, asghari2019deep, kim2022prediction, nguyen2025explainable}. For example, the characterization gap in the Tokyo-CLAY/14/67760 database \citep{otake2024exploring} has been examined using hierarchical Bayesian models, gradient-boosted tree ensembles, and Transformer-based deep learning, which are reported to outperform traditional empirical methods
\citep{otake2025geoai, manandhar2026predicting}. These models satisfy requirement~(a), but their accuracy depends on the representativeness of the training data and deteriorates outside the training domain \citep{wadoux2020machine} and they require about 80\% of the dataset to be used for training \citep{kim2022prediction,otake2025geoai,manandhar2026predicting}. Also, although physical consistency may be implicitly reflected in the training data, no mechanical laws explicitly constrain the predictions. Thus, the predictions are not guaranteed to be physically admissible beyond the training domain and rely on large training datasets, thereby not fulfilling requirements~(b) and~(c), respectively.

\subsubsection{Direct Physics-Based Modeling}
\label{sec:physicsbased}

A second approach calibrates a constitutive model against measured soil responses and then uses the calibrated model to reproduce the required mechanical properties. Models such as Cam-Clay \citep{roscoe1968generalised} and the Hardening Soil model \citep{schanz1999hardening} represent strength, stiffness, compressibility, and yielding through a finite set of constitutive parameters \citep{kadlicek2022automated}. These parameters are calibrated using the experimental results of odometer and triaxial tests which provide the stress-strain responses and compression curves required for the same \citep{astm_d2435, astm_d4767}. Because the predicted response is constrained by the governing constitutive equations, this approach addresses requirement~(b) within the assumptions of the selected model. However, the basic properties play no role in this identification. Although correlations between the constitutive parameters and the plasticity index have been proposed \citep{wroth1978correlation, nakase1988constitutive, wood1990}, they are derived from reconstituted clays and thus characterize the intrinsic
rather than the natural behavior, which is additionally governed by the structure and stress history that is affected by the constitution process \citep{burland1990, leroueil1990general}. Consequently, requirement~(a) is not satisfied in these direct physics based modeling approaches.

\subsubsection{Hybrid and Physics-Informed Approaches}
\label{sec:hybrid}

A third category combines elements of the two preceding approaches. Hybrid approaches combine data-driven methods with mechanical models to improve the calibration efficiency, physical consistency, or both \citep{yuan2025physics, wang2024digital}, and may be divided into two forms according to the role played by the mechanical model. The first form is an automated version of the direct physics approach where a constitutive model is used to represent the geomaterial response, while an optimization algorithm or machine learning is used to calibrate the constitutive model parameters \citep{calvello2004selecting, machacek2022automatic}. Approaches like Finite element model updating and related inverse-analysis methods adjust these parameters until the simulated response reproduces measured stress-strain curves, settlements, or field displacements \citep{tariq2025inverse, ghorbani2024parameters}. Since a physically meaningful constitutive model is used, these methods satisfy requirement~(b). However, a direct prediction from basic properties to mechanical properties is not obtained and thus requirement (a) is not satisfied.  In the second form, physics-informed neural networks represent the constitutive
behavior directly or are coupled with finite element analysis
\citep{liu2026fempinn, wang2024tensor, guan2023finite}. Mechanical constraints like 
material frame indifference \citep{heider2020so3}, strain-rate independence
\citep{zhang2020using, weber2023physically}, enforcement of the equilibrium
equations \citep{raissi2019physics, haghighat2021physics}, and non-negative
energy dissipation \citep{irani2025thermodynamically, he2022thermodynamically} are enforced within the architecture. Due to these constraints requirement~(b) is satisfied. However, as the training data used are experimental measurements, full-field observations, or synthetic constitutive responses rather than routine basic properties, requirement (a) and (c) may not be simultaneously satisfied. Thus, neither form
provides a direct prediction pathway that begins from routine basic properties and ends with the mechanical properties.

\subsection{Research Gap and Proposed Framework}
\label{sec:gap}

To address the characterization gap while satisfying requirements~(a), (b),
and~(c) identified above, we propose BRIDGE, \textbf{B}ridging \textbf{R}outine \textbf{I}nputs and \textbf{D}igital modelling for \textbf{G}eotechnical \textbf{E}ngineering, a modular framework that predicts mechanical properties directly
from routinely available basic properties. 

The dataset is first screened to exclude outliers and non-physical responses,
after which it is partitioned into Cells on the basis of the basic properties
and $S_u$. A sampling method then selects a calibration dataset of no more than
20\% of the samples in each Cell, while the remaining samples form the
validation dataset. In the present work, we adopt a finite element
implementation of the Extended Cam-Clay (ECC) constitutive model as the
mechanical model, whose constitutive parameters are identified for every
calibration sample through a multi-stage optimization process. Only converged
calibrations are retained, and the ECC models so obtained collectively define
the mechanical surrogate of the Cell, from which the admissible ranges of the
ECC parameters and the corresponding predictive envelope in the
mechanical property space are obtained. Each validation sample is subsequently
assessed against the predictive envelope of its own Cell, and a Cell is regarded
as validated when a majority of its validation samples fall within this
envelope. Once a Cell has been validated, its mechanical surrogate can be
applied to any new sample whose basic properties lie within the basic property limits of that Cell, so that the mechanical property ranges are predicted
directly from the basic properties.

BRIDGE therefore satisfies the three requirements, namely (a) prediction from
routinely available basic properties to mechanical properties, (b) physical consistency enforced through the calibrated mechanical surrogate, and (c) calibration from a limited number of samples. The framework is furthermore modular, since the source database,
mechanical model, calibration parameters, binning criterion, and
sampling method may each be independently changed without altering the underlying architecture.

\subsection{Paper Organization}
\label{sec:organization}

The remainder of the paper is organized as follows. Section~\ref{sec:overview} provides an overview of BRIDGE and its overall workflow. Section~\ref{sec:dataset} describes the marine clay database, and discusses the available binning and sub-binning options. Section~\ref{sec:workflow} details the preparation of the calibration and validation datasets through data screening, binning and sub-binning, and sampling. Section~\ref{sec:calibration} presents the multi-stage calibration procedure, including the forward model and staged calibration and Section~\ref{sec:envelope} describes the construction of the predictive envelope. Section~\ref{sec:demo} the entire workflow of the BRIDGE framework through a representative example while Section~\ref{sec:numerical} presents numerical studies that evaluate the effects of the binning criterion, second-level binning, sampling method, and sample budget on calibration reliability and accuracy. Finally, Section~\ref{sec:conclusions} summarizes the primary findings and identifies directions for future work.
%%%%%%%%%%%%%%%%%%%%%%%%%%%%%%%%%%%%%%%%%%%%%%%%%%%

%%%%%%%%%%%%%%%%%%%%%%%%%%%%%%%%%%%%%%%%%%%%%%%%%%%
\section{Overview of the BRIDGE Framework}
\label{sec:overview}

\begin{figure}[H]
  \centering
  \includegraphics[width=\textwidth, trim={0cm 1cm 0cm 1cm}, clip]{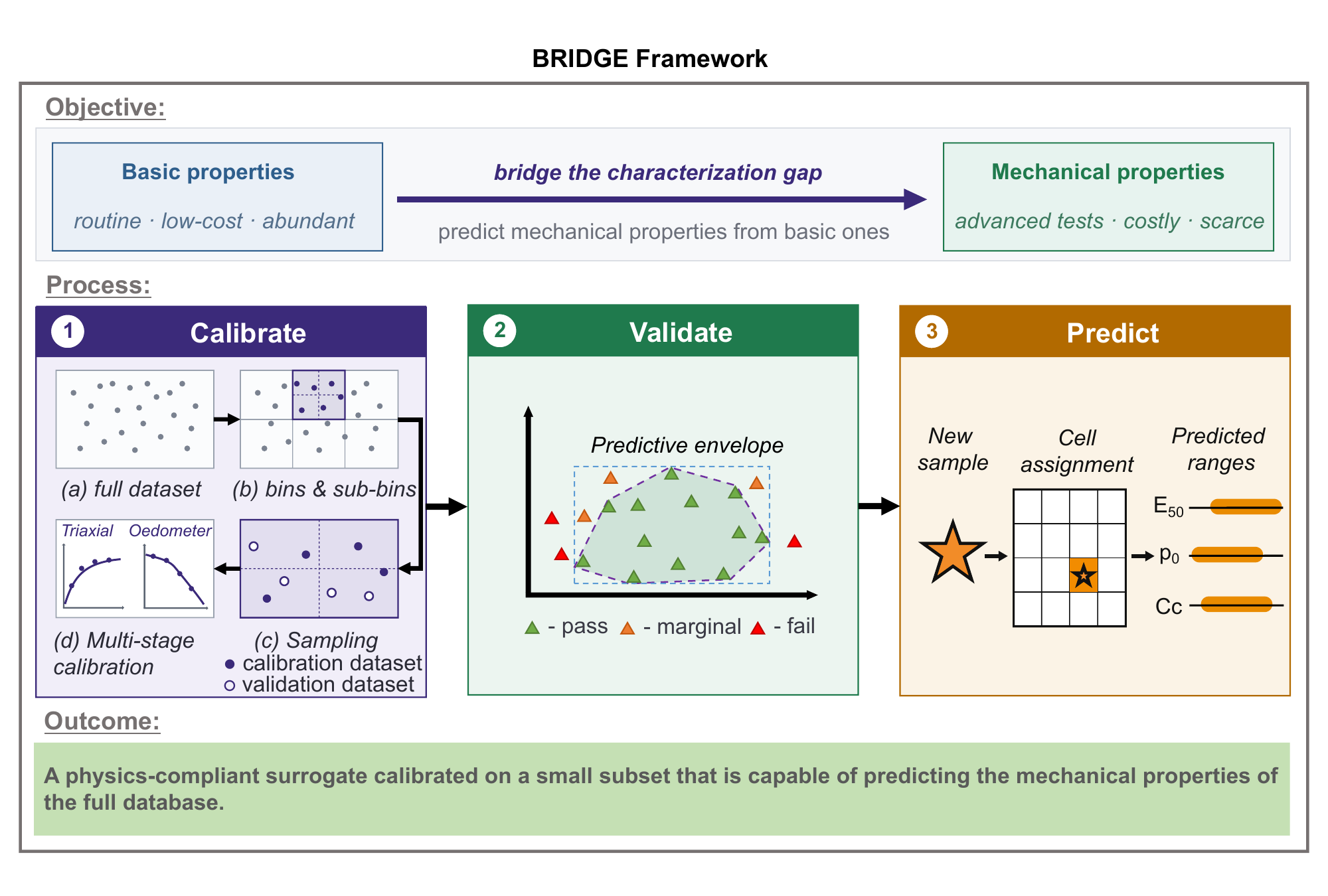}
  \caption{Conceptual overview of BRIDGE. The workflow bridges the characterization gap between basic and mechanical properties through calibration, validation, and prediction. Cells are formed through one or two level binning, and a small calibration dataset is used to identify ECC parameter ranges and construct a predictive envelope. Samples in the validation dataset are assessed against this envelope and classified as pass, marginal, or fail. Once a Cell has been validated, its mechanical property ranges can be used to predict $E_{50}$, $p_o$, and $C_c$ ranges for new samples whose basic properties and $S_u$ fall within the Cell limits.}
  \label{fig:BRIDGE_overview_simple}
\end{figure}

This section presents the components of BRIDGE in the order in which the
framework employs them. Fig.~\ref{fig:BRIDGE_overview_simple}
presents the conceptual overview of this framework that addresses the characterization gap between basic and mechanical properties. To this end, a mechanical model is calibrated on a small portion of the dataset and validated against the remaining samples through a predictive envelope. The outcome is a physics-compliant surrogate capable of predicting the mechanical properties of the entire dataset from the basic properties alone. Throughout, the framework operates on three categories
of soil properties, namely (a) the basic properties, (b) the constitutive parameters of the mechanical model, here the ECC constitutive model, and (c) the mechanical properties. 

The \textit{basic properties} are the routine classification measurements, namely the liquid limit $LL$, the plastic limit $PL$, the water content $w$, the void ratio $e$, the degree of saturation $S_{r}$, and the total unit weight $\gamma_{t}$. The \textit{mechanical properties} are the design quantities obtained from advanced laboratory tests, namely the undrained
shear strength $S_{u}$, the undrained secant stiffness $E_{50}$, the
preconsolidation pressure $p_{o}$, and the compression index $C_{c}$. Here we emphasize that $S_{u}$, although an mechanical property, can also be
estimated rapidly and inexpensively in the field with a pocket penetrometer or vane shear \citep{astm_d8121, astm_d8605, mayne2002subsurface}. The readings
of these devices correlate well with laboratory strength measurements, and
their scatter \citep{budak2022practical, okelly2023discussion} is consistent
with the documented variability of $S_{u}$ itself
\citep{yasun2018capability, mousavi2021modeling, phoon1999characterization}.
This allows $S_{u}$ to be considered for the binning process described in Section~\ref{sec:binning_rationale}.

The
\textit{constitutive parameters} link the basic and mechanical properties. They are the
inputs of the ECC model \citep{abaqus2024analysis} and govern its mechanical response. The framework organizes them into a calibration set (CS) and a fixed set (FS). The CS
comprises the logarithmic elastic bulk modulus $\kappa$, the logarithmic
plastic bulk modulus $\lambda$, the critical-state stress ratio $M$, and the
initial yield-surface size $a_{0}$. These four parameters are identified
through a multi-stage calibration process (Section~\ref{sec:calibration}). The FS comprises of the void ratio $e$, the density $\rho$,
the permeability $k$, and Poisson's ratio $\nu$.~\ref{apx:ecc} presents key details of the ECC constitutive model adopted here. 

\begin{figure}[H]
  \centering
  \includegraphics[angle=0, width=\textwidth, trim={12.5cm 1.5cm 10.5cm 6.5cm}, clip]{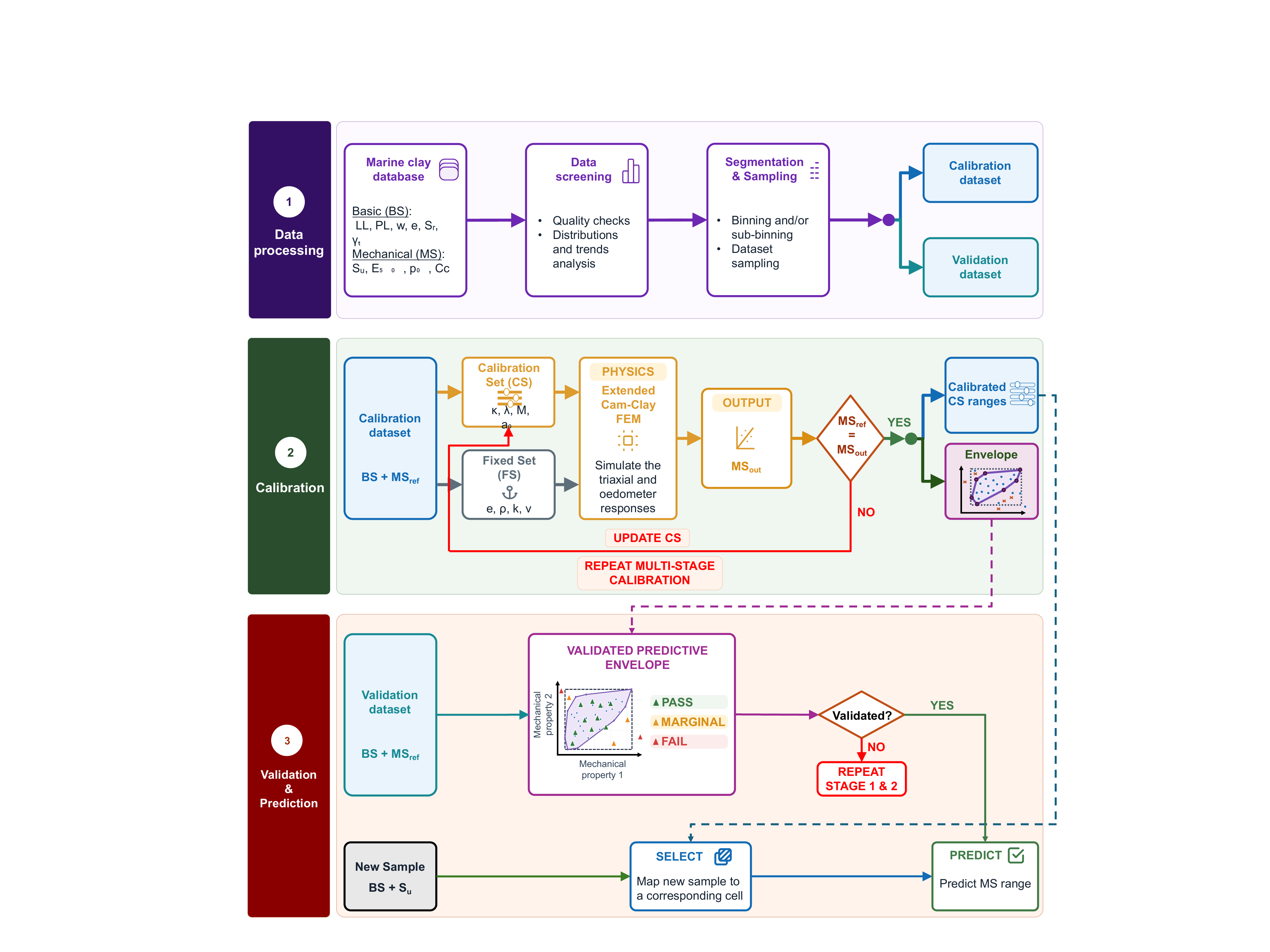}
  \caption{Detailed workflow of the BRIDGE framework in three stages. In Stage 1, the marine clay database comprising of the basic properties (BS: $LL$, $PL$, $w$, $e$, $S_r$, $\gamma_t$) and mechanical properties (MS: $S_u$, $E_{50}$, $p_o$, $C_c$) is screened through quality checks and analyses of its trends. Binning and optional two-level binning partition the screened dataset into Cells, and sampling forms the Cell-specific calibration and validation datasets. In Stage 2, the reference mechanical properties ($\mathrm{MS}_{\mathrm{ref}}$) of each calibration sample drive the multi-stage calibration of the calibration set (CS: $\kappa$, $\lambda$, $M$, $a_0$). We assign the fixed set (FS: $e$, $\rho$, $k$, $\nu$) to each sample and hold it constant throughout the calibration. Repeated finite element (FE) simulations of the triaxial and oedometer responses use the ECC model adopted in this work and update the CS until the simulated mechanical properties ($\mathrm{MS}_{\mathrm{out}}$) reproduce $\mathrm{MS}_{\mathrm{ref}}$ within a prescribed tolerance. Only converged calibrations are retained to define the Cell-specific CS ranges and predictive envelope. Stage 3 begins with validation of the Stage 2 calibration. The measured mechanical properties of the validation samples are mapped directly onto the corresponding predictive envelope. We classify each sample as (a) pass if its mechanical property coordinates lie within the convex hull, (b) marginal if they lie outside the convex hull but within the bounding box, or (c) fail if they lie outside the bounding box. We consider a Cell validated when a majority of its validation samples are classified as pass or marginal. The basic properties and $S_u$ for a new sample are used to identify the appropriate Cell, which can subsequently be used to predict its $E_{50}$, $p_o$, and $C_c$ ranges.}
  \label{fig_BRIDGE_overview}
\end{figure}

 Fig.~\ref{fig_BRIDGE_overview}
details the complete workflow, which comprises of three sequential stages - data processing, calibration, and combined a validation and prediction stage. In Stage 1, the marine clay database described in Section~\ref{sec:dataset} is screened through quality checks and trend analyses (Section~\ref{sec:screening}). We then apply either one-level or two-level binning (Section~\ref{sec:binning}). One-level binning applies a single criterion to the entire dataset. Two-level binning applies a second sub-binning criterion within the bins defined by the first criterion. The smallest units produced by either procedure are referred to as Cells. Within each Cell, the sampling procedure described in Section~\ref{sec:sampling} selects a calibration dataset comprising no more than 20\% of the samples. The remaining samples form the validation dataset. Both sets retain all available basic properties and reference mechanical properties from the database.

In Stage 2, a multi-stage calibration (Section~\ref{sec:calibration})is performed for each sample of the calibration dataset. The analysis identifies the CS parameters through repeated finite element simulations of the triaxial and oedometer tests. One CS parameter is adjusted at a time until the simulated response matches the corresponding target mechanical property. For each Cell, the successfully calibrated samples define the calibrated CS ranges and predictive envelope in mechanical property space (Section~\ref{sec:envelope}). Thus, after calibration in each Cell, we can link the ranges of the basic properties and $S_u$ to a calibrated mechanical surrogate.

In Stage 3, we validate the Cell-specific calibration completed in Stage 2 to ensure that the mechanical surrogate can predict the mechanical properties of a new sample from its basic properties. For each Cell, we map the measured mechanical properties of the validation samples from that Cell onto the predictive envelope constructed at the end of Stage 2. A Cell is considered successfully validated when its predictive envelope represents the majority of the corresponding validation samples. The validated mechanical surrogate is then ready to predict the mechanical properties of any new sample whose basic properties place it within the same Cell without additional advanced laboratory tests.

Here, it is important to define the criterion by which the predictions of the validated Cells are assessed. As shown in Section~\ref{sec:dataset}, the basic and mechanical properties are weakly correlated, which limits what any approach addressing the characterization gap can deliver on this dataset. The objective of the prediction stage is therefore to assess the consistency of the BRIDGE framework, which uses less than 20\% of the samples of a Cell to predict the range of mechanical properties of the remaining samples. We regard the framework as consistent when a predicted range of an mechanical property that encloses a given fraction of the calibration samples also encloses at least approximately the same fraction of the samples used in prediction. For example, the 50\% prediction range, defined as the interval that encloses half of the calibrated values of an mechanical property in a Cell, should also contain about half of that of the prediction samples. Similarly, the 80\% prediction range, defined analogously, should cover the measured mechanical property values of about four in five prediction samples. Since the width of these predicted ranges reflects the natural scatter of the measured data within each Cell rather than a limitation of the BRIDGE framework, narrowing them is not an objective of the present work and is left for future development.

Finally, we emphasize that BRIDGE is modular. The source database, the binning criteria and level, the sampling method, the mechanical model, and the composition of the CS and the FS can each be adapted to the application at hand without altering the overall architecture. This modularity allows BRIDGE to be applied to other databases and geotechnical applications while preserving the sequence of binning, calibration, validation, and prediction.

%%%%%%%%%%%%%%%%%%%%%%%%%%%%%%%%%%%%%%%%%%%%%%%%%%%
\section{Marine Clay Dataset description}
\label{sec:dataset}

\subsection{Dataset overview}
\label{sec:dataset_overview}

\begin{figure}[H]
\centering
\includegraphics[angle=0, width=1\textwidth,trim = 0cm 0cm 0cm 0cm, clip]{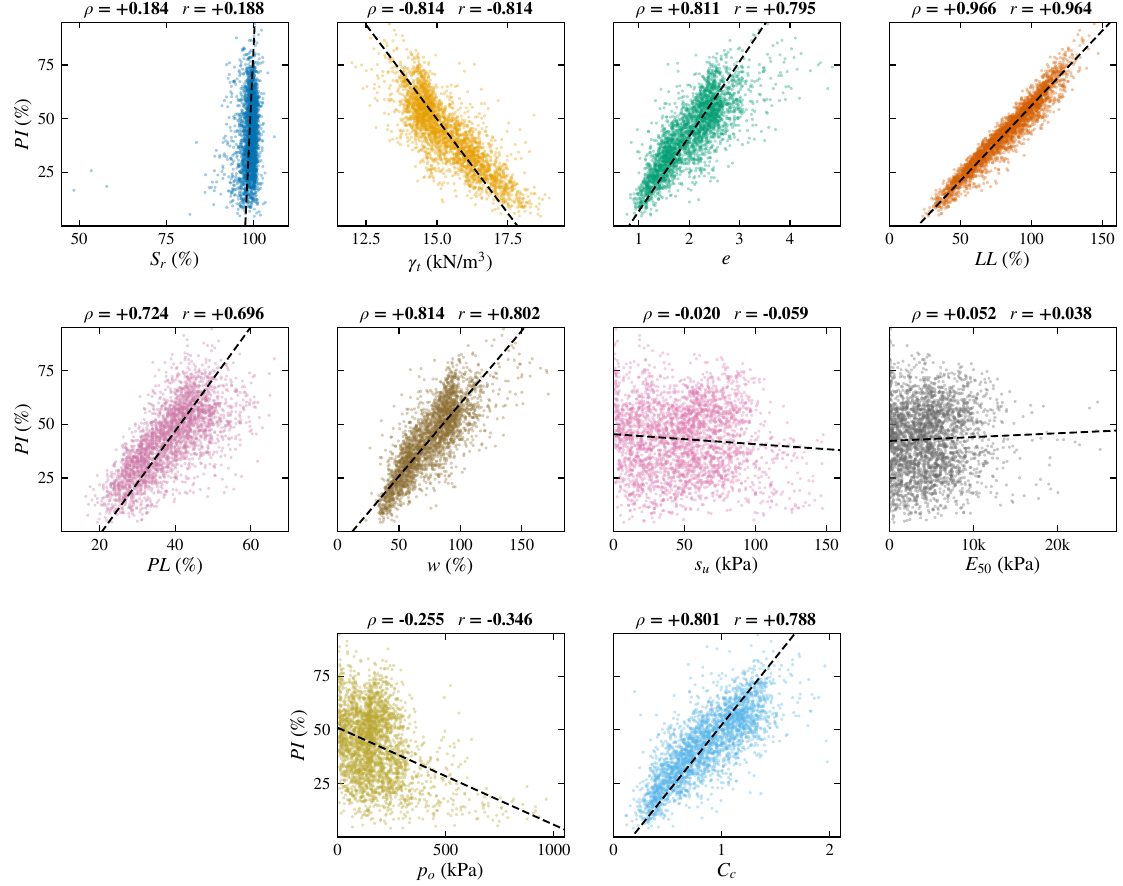}
\caption{Crossplots between the plasticity index and each remaining recorded parameter of the working dataset (2,766 samples). $PI$ is computed as $PI = LL - PL$. Each panel reports the Spearman rank correlation $\rho$ and the Pearson correlation $r$. The dashed lines represent linear fits.}
\label{fig_parameter_relations}
\end{figure}

This subsection reviews and analyzes the marine clay dataset used to develop and evaluate BRIDGE. The working dataset is drawn from the Tokyo-CLAY database \citep{otake2024exploring} and contains 2,766 samples with 10 recorded properties. Each sample includes both routinely measured basic properties and laboratory-measured mechanical properties. Table~\ref{tab:dataset} lists these parameters together with their symbols, units, and summary statistics. Six of the values are basic properties, while the remaining four, namely the undrained shear strength $S_u$, the undrained secant modulus $E_{50}$, the preconsolidation pressure $p_o$, and the compression index $C_c$, are mechanical properties that are conventionally measured on undisturbed samples through oedometer and triaxial tests. The database does not directly report the plasticity index ($PI$). We calculate it from the liquid limit $LL$ and plastic limit $PL$ as $PI = LL - PL$. This derived basic property is included in the correlation analysis presented in Fig.~\ref{fig_parameter_relations}.
 
\begin{table}[htbp]
    \caption{Parameters of the working marine clay dataset (2,766 samples): symbols, units, tier assignment, and summary statistics.}
    \label{tab:dataset}
    \centering
    \begin{tabular}{llllrrrr}
       \toprule
        Parameter & Symbol & Unit & Group & Min & Mean & Max & $\sigma$ \\
        \midrule
        Liquid limit & $LL$ & - & Basic & 24.0 & 81.5 & 146.4 & 22.9 \\
        Plastic limit & $PL$ & - & Basic & 16.4 & 38.4 & 66.1 & 8.4 \\
        Natural water content & $w$ & - & Basic & 23.0 & 75.5 & 171.5 & 22.2 \\
        Void ratio & $e$ & - & Basic & 0.76 & 2.04 & 4.83 & 0.58 \\
        Degree of saturation & $S_r$ & - & Basic & 48.6 & 98.8 & 105.8 & 2.4 \\
        Total unit weight & $\gamma_t$ & kN/m$^3$ & Basic & 12.0 & 15.4 & 18.9 & 1.1 \\
        Undrained shear strength & $S_u$ & kPa & Mechanical & 0.7 & 49.5 & 147.1 & 29.4 \\
        Undrained secant modulus & $E_{50}$ & kPa & Mechanical & 1 & 4693 & 24946 & 3465 \\
        Preconsolidation pressure & $p_o$ & kPa & Mechanical & 6.9 & 175.9 & 961.1 & 125.7 \\
        Compression index & $C_c$ & - & Mechanical & 0.11 & 0.85 & 1.97 & 0.33 \\
        \bottomrule
    \end{tabular}
\end{table}
 
Fig.~\ref{fig_parameter_relations} presents crossplots of $PI$ against each of the remaining recorded parameters, in which each panel reports the Spearman rank correlation coefficient $\rho$ \citep{spearman1904} and the Pearson correlation coefficient $r$ \citep{pearson1895} together with a linear fit. The definitions and expressions for both correlations are reported in ~\ref{apx:correlation_equations}. Both coefficients are bounded between $-1$ and $1$, where the bounds correspond to perfect negative and perfect positive association, respectively. Magnitudes close to one indicate a strong association, magnitudes close to zero indicate a weak association, and the sign indicates whether one parameter increases or decreases with the other. Similar nonzero values of $\rho$ and $r$ suggest an approximately linear relationship between the parameters. A substantial difference between the two, typically with $|\rho|$
exceeding $|r|$, indicates a monotonic yet nonlinear relation between the parameters.

Based on these correlations, two observations can be drawn from Fig.~\ref{fig_parameter_relations}. First, $PI$ is strongly correlated with the basic properties, with $\rho = +0.97$ against $LL$, $\rho = +0.81$ against both $w$ and $e$, and $\rho = -0.81$ against $\gamma_t$, where the correlation with $LL$ is partly by construction since $PI$ is derived from the Atterberg limits. Second, the association between $PI$ and the mechanical properties is not uniform. $C_c$ does relate to $PI$ closely, with $\rho = +0.801$, whereas $S_u$, $E_{50}$, and $p_o$ are nearly uncorrelated with $PI$, with $\rho$ values of $-0.02$, $+0.05$, and $-0.26$, respectively. Consequently, samples with nearly identical $PI$ values span almost the entire observed range of $S_u$, $E_{50}$, and $p_o$. These trends indicate that the basic properties are sufficient to characterize the $C_c$ of these clays but are unable to represent the mechanical properties governed by the stress history or strength of the sample. This is a quantitative representation of the characterization gap in this dataset.
 
\begin{figure}[H]
\centering
\includegraphics[angle=0, width=1\textwidth,trim = 0cm 0cm 0cm 0cm, clip]{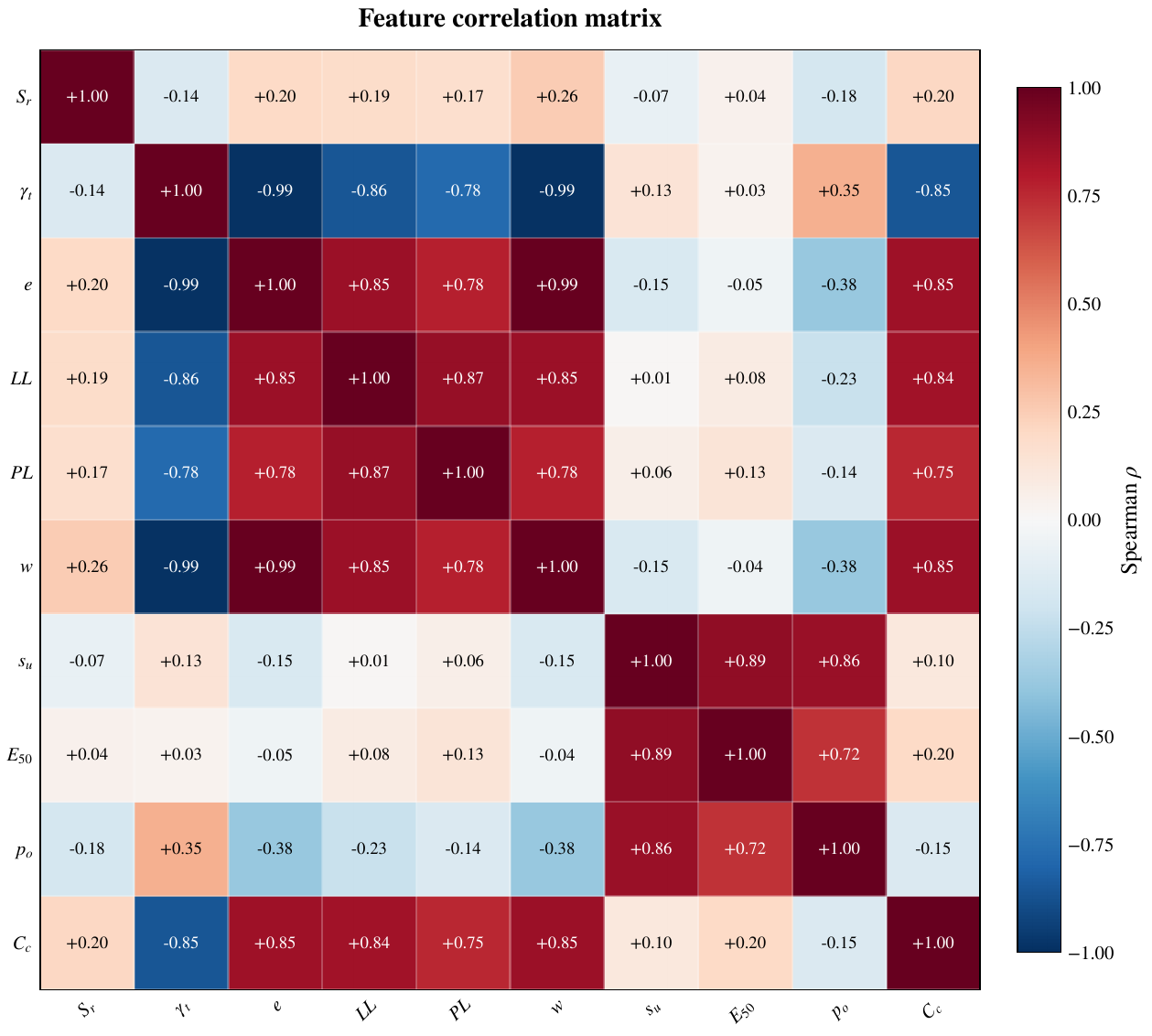}
\caption{Spearman rank correlation matrix of the ten recorded parameters of the working dataset. Each tile reports the coefficient $\rho$ for one parameter pair.}
\label{fig_Parameter_heatmap_spearman}
\end{figure}

Fig.~\ref{fig_Parameter_heatmap_spearman} extends this pairwise examination to all parameter combinations by reporting the Spearman rank correlation coefficient for every parameter pair. The correlation matrix reveals that the recorded parameters can be organized into two blocks. The first block comprises the basic properties, in which the void ratio and the natural water content are almost perfectly correlated ($\rho = +0.99$), both are strongly negatively correlated with the total unit weight ($\rho = -0.99$), and the Atterberg limits follow the same cluster with $\rho$ values between $+0.75$ and $+0.87$. $C_c$ is also associated with this block, with $\rho$ values of $+0.84$ to $+0.85$ against $LL$, $w$, and $e$. The second block comprises $S_u$, $E_{50}$, and $p_o$, which are strongly correlated with one another, with pairwise $\rho$ values of $+0.89$, $+0.86$, and $+0.72$, while remaining weakly correlated with all members of the first block ($|\rho| \leq 0.38$). 
The correlation matrix informs the design of the BRIDGE framework, as the mechanical property block contains its prediction targets. For the calibration to be successful, the dataset must be segmented into bins of samples with similar mechanical responses. However, the basic properties correlate only weakly with $S_{u}$, $E_{50}$, and $p_{o}$. Thus, binning on the basic properties alone is unlikely to be sufficient. Thus, we consider and quantitatively compare different binning options in Section~\ref{sec:binning_rationale}.

\subsection{Dataset binning options}
\label{sec:binning_rationale}
 
\begin{figure}[H]
\centering
\includegraphics[width=1\textwidth,trim = 2cm 4cm 2cm 4.5cm, clip]{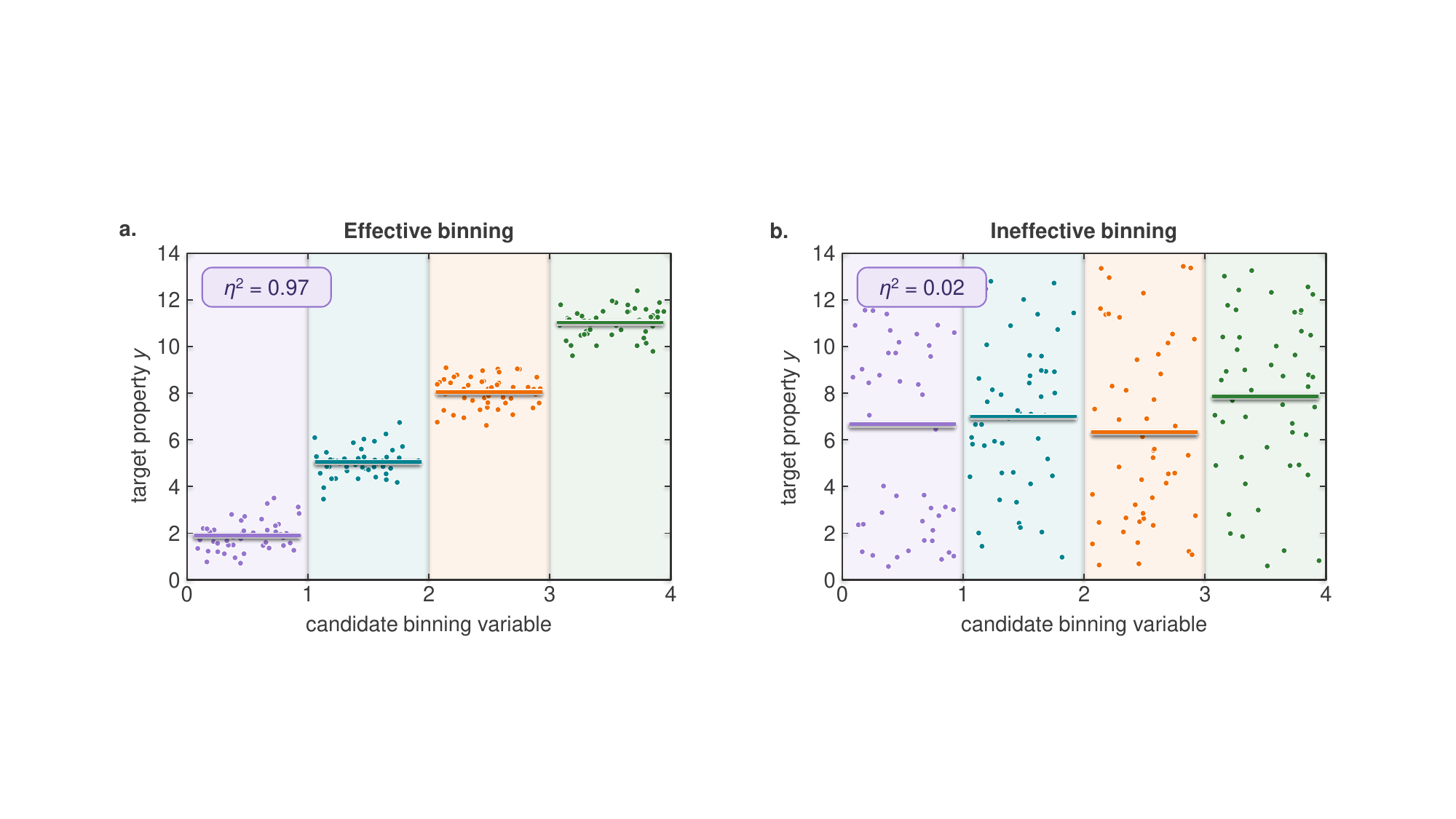}
\caption{Schematic representation of the concept of $\eta^{2}$, illustrated with synthetic data partitioned into four bins along a candidate binning variable. (a) A binning in which the target property $y$ scatters little within each Bin and the Bin means are well separated attains $\eta^{2}=0.97$. (b) A binning in which every Bin reproduces the full scatter of the target and the Bin means nearly coincide has $\eta^{2}=0.02$. The horizontal lines denote the Bin means, and the shaded bands denote the bins.}
\label{fig:eta2_concept}
\end{figure}

The BRIDGE framework is meaningful only when it is applied to bins of samples with comparable property ranges. This requires each Bin to be defined such that a few of its samples remain representative of the Bin as a whole. An effective binning criterion must therefore satisfy two requirements, namely (a) minimal within-Bin scatter and (b) well-separated Bin means. Ranking the candidate binning variables against these requirements calls for a single objective metric that captures both simultaneously. To this end, we adopt the correlation ratio $\eta$ \citep{pearson1905general}. Fig.~\ref{fig:eta2_concept} presents this concept on synthetic data, in which a binning criterion that satisfies both requirements attains a value of $\eta^2$ close to one while one that fails both attains a value close to zero. For a candidate binning variable divided into $K$ bins and a target property $y$,
\begin{equation}
\eta^2
= \frac{\mathrm{SS}_{\mathrm{between}}}{\mathrm{SS}_{\mathrm{total}}}
= \frac{\sum_{k=1}^{K} n_k \left(\bar{y}_k - \bar{y}\right)^2}
       {\sum_{i=1}^{N} \left(y_i - \bar{y}\right)^2},
\label{eq:eta2}
\end{equation}
where $y_i$ is the target value of sample $i$, $\bar{y}$ is the overall mean over the $N$ samples, and $n_k$ and $\bar{y}_k$ are the count and the mean of Bin $k$.  The correlation ratio is dimensionless and bounded between zero and one, so that targets measured in different units can be compared directly. Higher values indicate better binning variables, since $\eta^{2}$ increases with decreasing within-Bin scatter and increasing separation of the Bin means. Any variable used to Bin the dataset should therefore attain a high $\eta^{2}$ value.

Five candidates, namely $w/LL$, $w/PL$, and $w/PI$, the void ratio $e$, and the undrained shear strength $S_u$, were evaluated against the four mechanical properties using $K = 8$ bins under both the \textbf{equally spaced (ES)} and the \textbf{equal-count (EC)} partitioning schemes. The ratio of water content and Atterberg limits have long been used to estimate the undrained shear strength \citep{wroth1978,wood1990,vardanega2014,kuriakose2017,shimobe2020}. The void ratio $e$ is the classical basis for estimating the compression index \citep{azzouz1976,burland1990} while the undrained shear strength $S_{u}$ is a standard variable for classifying soil behavior \citep{iso14688,ladd1974}. 

Under the \textbf{ES} scheme the range of each candidate is divided into eight intervals of identical width, whereas under the \textbf{EC} scheme the Bin edges are placed such that every Bin holds approximately equal number of samples. In the \textbf{ES} scheme, the outermost bins contained as few as one or two samples, rendering them unsuitable for calibration. The \textbf{EC} scheme is therefore adopted in this work. Table~\ref{tab:eta2} compiles the resulting values of $\eta^2$ for both schemes side by side. The $S_u$ is the strongest criterion for three of the four targets, attaining $\eta^2 = 0.67$ for $E_{50}$ and $0.53$ for $p_o$, whereas only $e$ explains the compression index best with $\eta^2=0.70$. In view of the above, we adopt $S_{u}$ as the binning variable and $e$ as the sub-binning variable within each $S_{u}$ Bin, thereby capturing the variability in $C_{c}$ that $S_{u}$ alone cannot resolve. Although $S_{u}$ is itself an mechanical property, its use for binning is admissible as it can be obtained during routine geomaterial characterization, as discussed in Section~\ref{sec:gap}. The $S_u$ binning and subsequent $e$ sub-binning together constitute the two-level binning procedure detailed in Section~\ref{sec:binning}. This choice of binning variables is examined numerically within the full calibration pipeline in Section~\ref{sec:studyA}. 
 
\begin{table}[H]
\centering
\caption{Squared correlation ratio $\eta^{2}$ of each mechanical property
against the five candidate binning variables, evaluated on the 2{,}766-sample
database with $K=8$ bins under the two partitioning schemes. The strongest
candidate for each target is marked in bold.}
\label{tab:eta2}
\begin{subtable}[t]{0.48\textwidth}
\centering
\caption{Equal-count (\textbf{EC}) [adopted in this work]}
\label{tab:eta2_ec}
\begin{tabular}{@{}lccccc@{}}
\toprule
Target & $w/LL$ & $w/PL$ & $w/PI$ & $e$ & $S_u$ \\
\midrule
$E_{50}$ & 0.09          & 0.08 & 0.03 & 0.08          & \textbf{0.67} \\
$S_u$    & 0.15          & 0.15 & 0.04 & 0.11          & \textbf{0.95} \\
$p_o$    & 0.06          & 0.23 & 0.03 & 0.22          & \textbf{0.53} \\
$C_c$    & $<\!0.01$     & 0.24 & 0.11 & \textbf{0.70} & 0.04          \\
\bottomrule
\end{tabular}
\end{subtable}
\hfill
\begin{subtable}[t]{0.48\textwidth}
\centering
\caption{Equally spaced (\textbf{ES})}
\label{tab:eta2_es}
\begin{tabular}{@{}lccccc@{}}
\toprule
Target & $w/LL$ & $w/PL$ & $w/PI$ & $e$ & $S_u$ \\
\midrule
$E_{50}$ & 0.06      & 0.08 & 0.03 & 0.08          & \textbf{0.66} \\
$S_u$    & 0.11      & 0.15 & 0.03 & 0.11          & \textbf{0.97} \\
$p_o$    & 0.04      & 0.23 & 0.06 & 0.22          & \textbf{0.58} \\
$C_c$    & $<\!0.01$ & 0.23 & 0.10 & \textbf{0.68} & 0.05          \\
\bottomrule
\end{tabular}
\end{subtable}
\end{table}
 
%%%%%%%%%%%%%%%%%%%%%%%%%%%%%%%%%%%%%%%%%%%%%%%%%%%
\section{Preparation of Calibration and Validation Datasets}
\label{sec:workflow}
 
This section presents the workflow through which the calibration and validation datasets are prepared from the marine clay database, corresponding to Stage 1 of Fig.~\ref{fig_BRIDGE_overview}. The workflow comprises three consecutive operations, namely (a) data screening, which enforces quality checks, (b) binning, one-level or two-level, to partition the retained samples into Cells with comparable property ranges, and (c)  sampling, which draws from each Cell a small calibration dataset and reserves the remaining samples as the validation dataset. The two datasets of each Cell subsequently serve as the inputs to Stages 2 and 3 of Fig.~\ref{fig_BRIDGE_overview}, respectively.

 \subsection{Data screening}
 \label{sec:screening}
In the first stage, we screen the 2,766 samples of the working dataset for physical admissibility through five filtering criteria that are applied in sequence. These are (a) $0.5 \leq e \leq 3.0$, (b) $LL > PL$, (c) $S_r \leq 105\%$, (d) $S_{u}$ $>$ 0;  $E_{50}$ $>$ 0, and (e) $E_{50}$ $\geq$ $S_{u}$. Applying the five filtering criteria removes 123, none, 1, none, and 33 samples, respectively, which leaves 2,609 samples. Samples violating any of these rules are excluded, since such violations can be attributed to measurement or recording error rather than to genuine soil behavior. Of the retained samples, 509 have reported degrees of saturation marginally above \(100\%\), with a maximum of \(103.1\%\). Because \(S_r\) is calculated from several measured quantities, small measurement errors can produce values slightly above \(100\%\), as observed experimentally \citep{peron2007improved,elkady2015stress}. A further 626 samples have natural water contents above their liquid limits and hence liquidity indices greater than unity. Such values are physically plausible for intact soft marine clays and have been documented in Ariake and Norwegian sensitive marine-clay deposits \citep{tanaka2001characterization,lheureux2019tiller}. Both groups were therefore retained.

\subsection{Binning and sub-binning}
\label{sec:binning}

\begin{figure}[htbp]
\centering
\includegraphics[width=1\textwidth,trim = 0.9cm 3cm 0cm 2.85cm, clip]{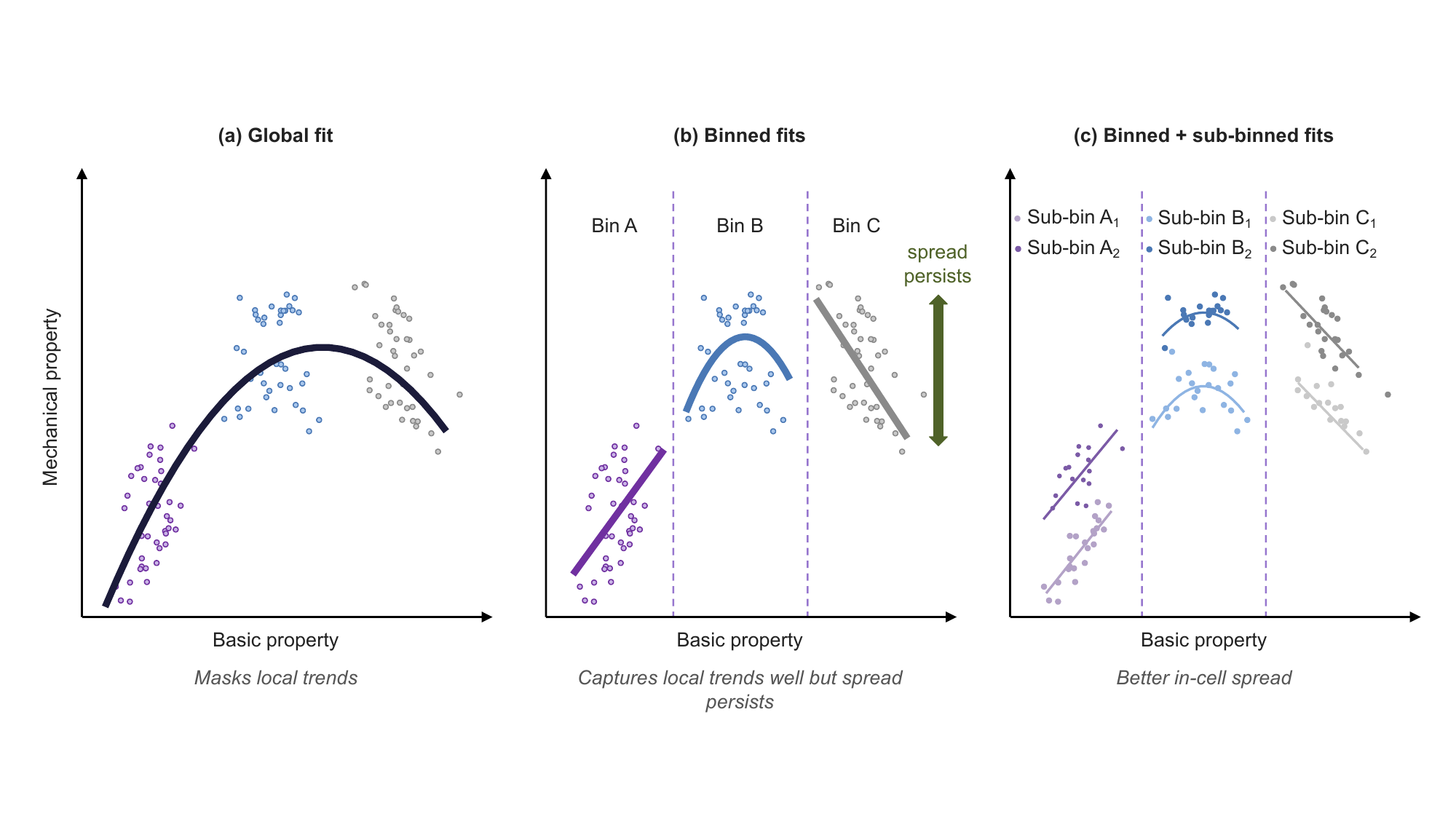}
\caption{Schematic representation, with synthetic data, of the two-level binning. (a) A single global fit across the full dataset masks the local trends. (b) Binning on the primary binning criteria ($S_u$) localizes the fits, while a wide spread could still persist within each Bin. (c) Sub-binning using a second variable $e$, indicated by the marker shading, further reduces the in-Cell spread.}
\label{fig:grouping}
\end{figure}

We partition the retained samples using two physically motivated variables,
namely the undrained shear strength $S_{u}$ and the void ratio $e$, which are
selected on the basis of the variance of each target property relative to the
remaining properties (Section~\ref{sec:binning_rationale}), whereby $S_{u}$
serves as the primary binning variable and $e$ as the secondary binning
(sub-binning) variable, as illustrated schematically in
Fig.~\ref{fig:grouping}. The Bin and Sub-Bin edges are placed as per the \textbf{EC} scheme. The smallest resulting unit is referred to as a Cell. The effect of binning and sub-binning are explored in Section~\ref{sec:numerical}.
 
\subsection{Sampling}
\label{sec:sampling}
The sampling operation completes the data processing stage by selecting a small calibration dataset from each Cell and reserving the remaining samples for validation. Sampling influences both the computational cost and the predictive capability of BRIDGE. Each calibration dataset sample undergoes the multi-stage calibration described in Section~\ref{sec:calibration}, which requires repeated finite element analyses. To limit the computational cost, we restrict the calibration dataset to no more than $20\%$ of the samples in each Cell. Also, sampling determines how well the calibration dataset represents its parent Cell. Only the successfully calibrated samples contribute to the Cell-specific CS ranges and predictive envelope defined in Section~\ref{sec:envelope}. An unrepresentative selection would consequently skew the predictive envelope and negatively impact the subsequent validation.

In view of the above, we construct the calibration dataset of each Cell with a prescribed number of samples $n_{s}$ through one of five candidate sampling methods, namely Kennard-Stone \citep{kennard1969computer}, Puchwein \citep{puchwein1988selection}, Latin hypercube selection \citep{mckay1979comparison}, Sobol selection \citep{sobol1967distribution}, and conditioned Latin hypercube \citep{minasny2006conditioned}. The selection operates on the basic properties alone, so that none of the predicted mechanical properties influences the composition of the calibration dataset. The sample budget $n_{s}$ and the sampling method govern the computational cost and the representativeness of the calibration dataset. A systematic comparison of the five methods in terms of the resulting accuracy is presented in Section~\ref{sec:studyC}.

\section{Multi-Stage Calibration}
\label{sec:calibration}

This section describes the forward model and multi-stage calibration procedure, depicted in Stage 2 of Fig.~\ref{fig_BRIDGE_overview}, used to calibrate the four ECC constitutive parameters $\lambda$, $\kappa$, $M$, and $a_0$ for each sample in the calibration dataset. The calibration procedure is implemented in Python and interfaces with the ABAQUS for all forward simulations. The successfully calibrated samples then define the Cell-specific calibration set (CS) ranges and predictive envelope used to validate the mechanical surrogate and support predictions for new samples.

\subsection{Forward Model}
\label{sec:forward}

The forward model comprises of two single-element simulations in
ABAQUS/Standard  \citep{abaqus2024analysis}, one for each laboratory test from which the mechanical properties originate namely the undrained triaxial and oedometer test. Both tests are simulated using a single unit axisymmetric CAX8RP element which, in the absence of body forces and under uniform face loading, represents a material point response of the Extended Cam-Clay formulation. A multi-element discretized specimen model would reproduce essentially the same response at a far greater computational cost which would make the multi-stage calibration protocol of Section~\ref{sec:stages} unfeasible. 

Each finite element analysis uses an ABAQUS input-file template in which the material parameters are represented by placeholders. For each sample, the four fixed set (FS) parameters, namely the initial void ratio $e$, mass density $\rho$, permeability $k$, and Poisson's ratio $\nu$, are assigned at the start of the calibration procedure and remain constant throughout the analysis. By contrast, the four calibration set (CS) parameters, namely $\lambda$, $\kappa$, $M$, and $a_0$, are updated before each finite element simulation of the multi-stage calibration procedure. The geometry, boundary conditions, and solver settings remain unchanged across all samples. Both models use unsymmetric matrix storage and automatic time incrementation, with the pore pressure change per increment limited to
$4\times10^{4}$ and $4\times10^{8}\ \mathrm{Pa}$ and the time increment to
$1$ and $500\ \mathrm{s}$ for the triaxial and oedometer models,
respectively. Automatic stabilization is applied with a dissipated energy
fraction of $2\times10^{-4}$ and an adaptive accuracy tolerance of $0.05$.
 
\begin{figure}[H]
\centering
\includegraphics[width=1\textwidth,trim = 5cm 0.5cm 5cm 1.75cm, clip]{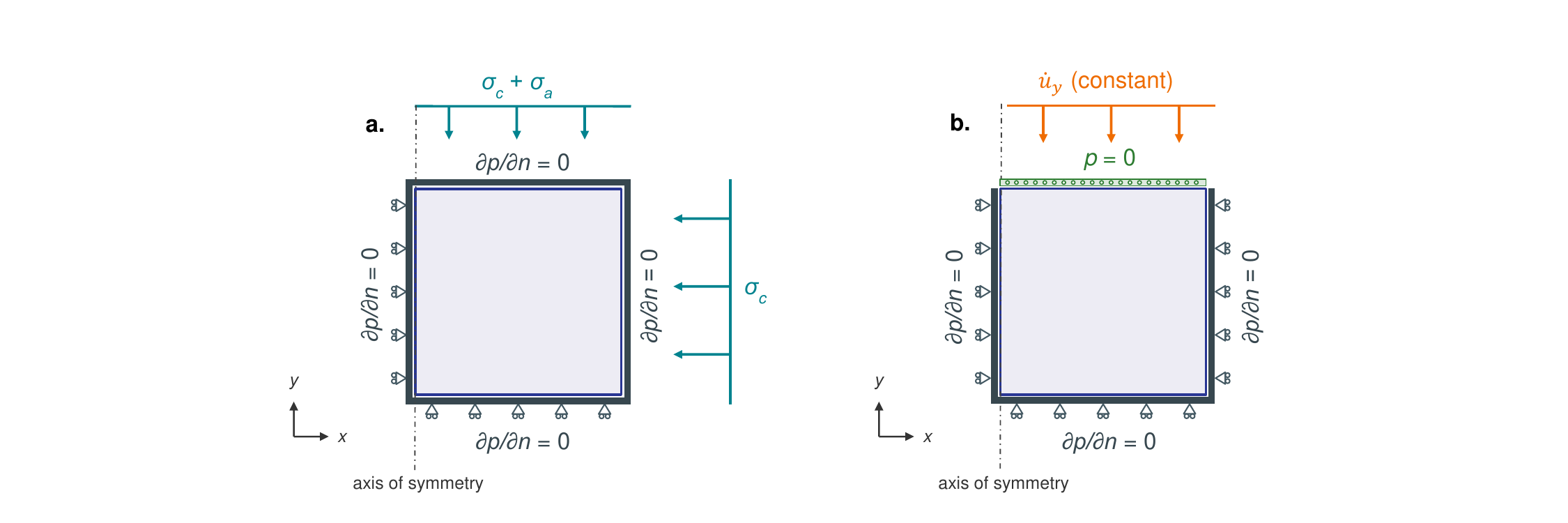}
\caption{Single-element axisymmetric models of the two simulated laboratory tests, where $p$ denotes the pore pressure. (a) Undrained triaxial model, in which the confining pressure $\sigma_c$ acts on the lateral and top faces, the axial stress $\sigma_a(t)$ is applied on the top face, and all boundaries are impermeable with $\partial p/\partial n = 0$. (b) Constant rate of strain oedometer model, in which the lateral boundary is radially restrained, the top face is loaded at a constant velocity $\dot{u}_y$ and drained at zero pore pressure, $p = 0$, while the remaining boundaries are impermeable with $\partial p/\partial n = 0$.}
\label{fig:single_element_models}
\end{figure}

\begin{figure}[H]
\centering
\includegraphics[width=1\textwidth,trim = 3cm 4cm 3cm 4cm, clip]{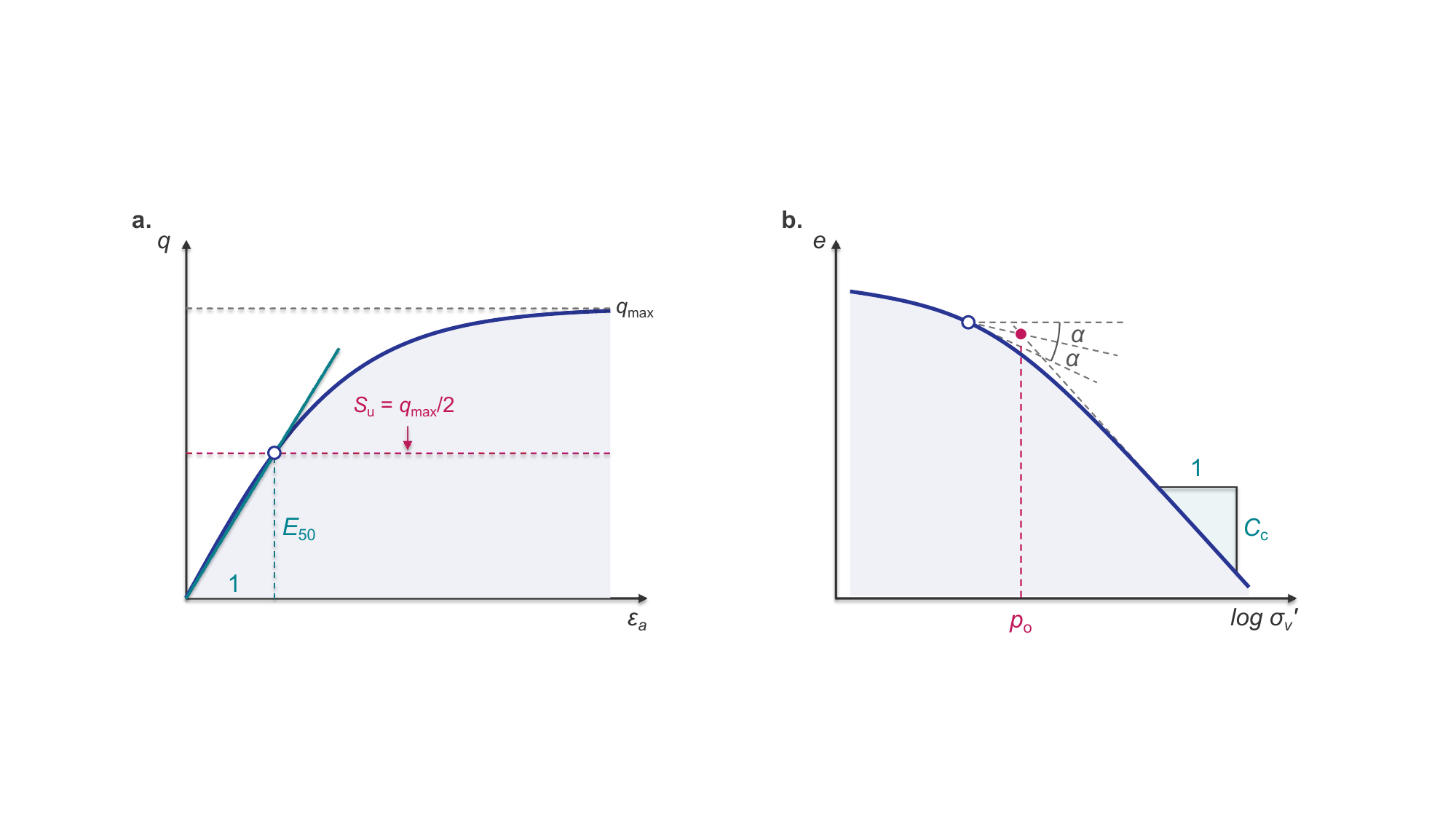}
\caption{Extraction of the mechanical properties from the simulated responses. (a) From the deviatoric stress $q$ against axial strain $\varepsilon_a$ response of the triaxial model, $S_u$ is half of the peak deviatoric stress $q_{\max}$ and $E_{50}$ is the slope of the secant drawn from the origin to the curve at $q_{\max}/2$. (b) From the compression curve of the oedometer model, the compression index $C_c$ is the slope of the virgin compression line and the preconsolidation pressure $p_o$ is extracted using the Casagrande method \citep{casagrande1936preconsolidation}.}
\label{fig:property_extraction}
\end{figure}

In the triaxial model, Fig.~\ref{fig:single_element_models}a, we reproduce a consolidated undrained test in two steps, in which a static step first applies pressure on the lateral and top faces, and a loading step then shears the specimen while keeping all boundaries impermeable. The total axial stress is ramped linearly by $380\ \mathrm{kPa}$ over the $100\ \mathrm{s}$ loading step. From the recorded deviatoric stress vs axial strain response, $S_u$ is measured as half of the peak deviatoric stress and $E_{50}$ as the  corresponding slope of the curve as shown in Fig.~\ref{fig:property_extraction}a. The deviatoric stress is defined as $q = \sigma_1 - \sigma_3$, where $\sigma_1$ and $\sigma_3$ denote the stresses in the vertical and horizontal directions, respectively.
 
In the oedometer model, Fig.~\ref{fig:single_element_models}b, the displacement of the lateral boundary is constrained. The displacement of the top face is ramped linearly to $0.25\ \mathrm{m}$ over $120{,}000\ \mathrm{s}$. This compresses the specimen to a nominal axial strain of 25 percent. The resulting strain rate of $2.13\times10^{-6}\ \mathrm{s^{-1}}$ corresponds to approximately 0.77 percent per hour. This rate lies within the range customary for soft clays  \citep{wissa1971,astm_d4186}. Drainage is permitted only through the loaded face, where the pore pressure $p_w$ is held at zero. The compression index $C_c$ and the preconsolidation pressure $p_o$ are extracted from the simulated compression curve, with $p_o$ identified by the Casagrande method \citep{casagrande1936preconsolidation} as illustrated in Fig.~\ref{fig:property_extraction}b.

\begin{figure}[H]
  \centering
  \includegraphics[width=1\textwidth,trim = 2cm 3cm 2cm 2.75cm, clip]{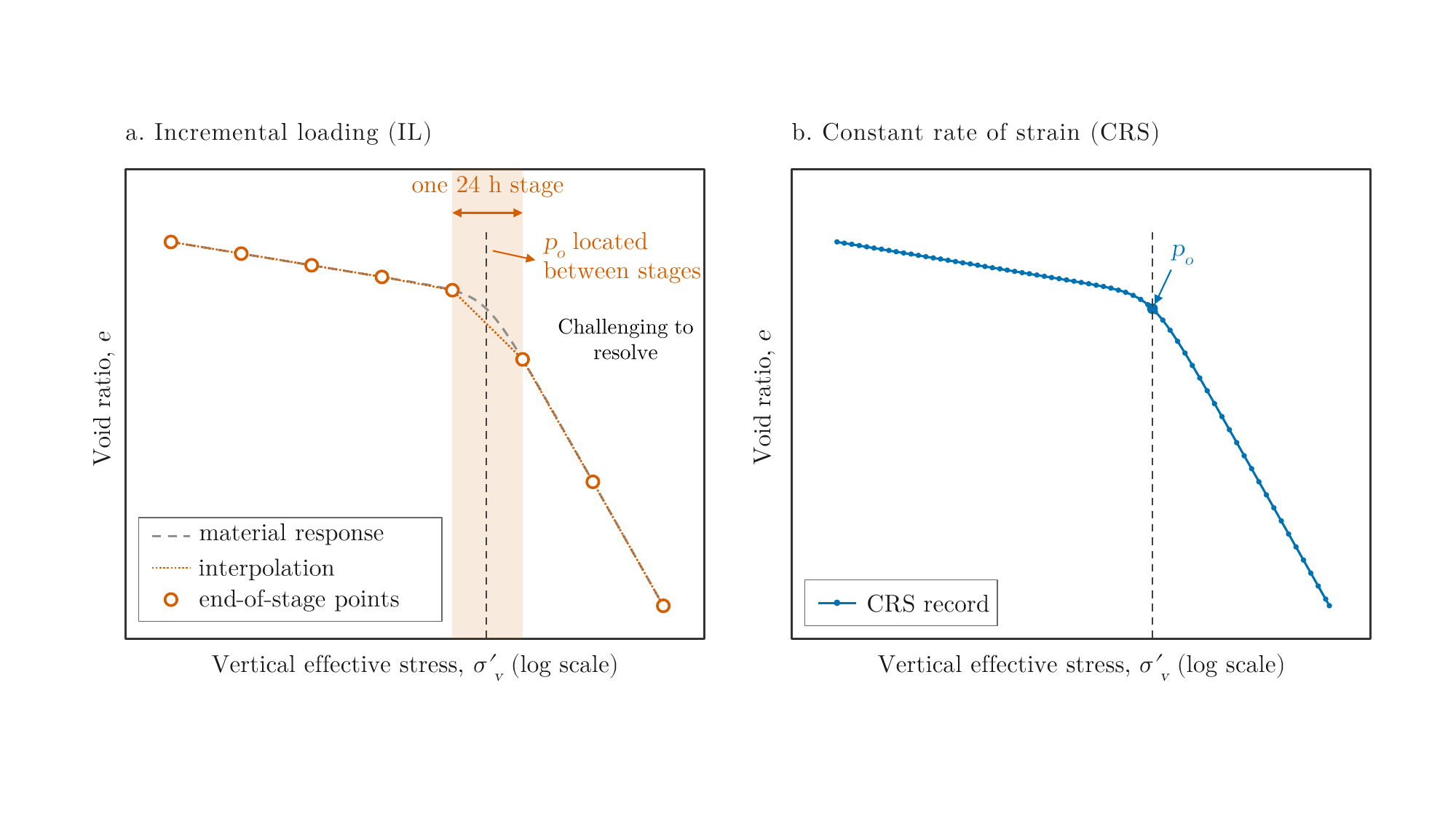}
  \caption{Schematic comparison of the resolution of the compression curve obtained from the two oedometer approaches, where both panels show the same material response. (a) The incremental loading approach records the response only at the end-of-stage points, whose spacing of one load doubling per 24 hr stage causes the preconsolidation pressure $p_o$ to fall in-between successive stages, which makes it challenging to identify accurately. (b) The CRS approach records the compression curve quasi-continuously, thereby resolving $p_o$ easily.}
  \label{fig:CRS_IL_resolution}
\end{figure}

The oedometer test can be modeled in either the constant rate of strain (CRS) configuration or the incremental loading (IL) configuration. In the IL configuration, the vertical stress
is applied in discrete stages, typically doubled at each stage and maintained for approximately 24 $hrs$ \citep{astm_d2435} per stage. In the CRS configuration, the specimen is compressed continuously at a prescribed axial strain rate, and the corresponding vertical stress response is recorded \citep{astm_d4186}. When comparing the two approaches, the CRS configuration has two distinct advantages. First, the CRS test traces the compression curve continuously, as illustrated schematically in Fig.~\ref{fig:CRS_IL_resolution}. This provides sufficient resolution to extract $C_{c}$ and $p_{o}$ as the CS is updated during calibration. The IL configuration by contrast produces only one point on the curve per 24 $hrs$ load stage. The resultant compression curve thus lacks the resolution to identify the unique values of $C_{c}$ and $p_{o}$ produced by successive CS updates. Second, the CRS test concludes in approximately 1.4 days of pseudo-time, whereas the IL configuration would require more than a week. Because the calibration stage requires thousands of finite element simulations, the higher process time of the IL configuration would impose a prohibitive computational cost. Considerable experimental work has compared CRS consolidation with conventional IL testing \citep{smith1969consolidation, wissa1971consolidation, sheahan1997experimental, lavorati2023evaluation}. These studies indicate that well selected CRS strain rates produce compression curves and estimates of $C_c$ and $p_o$ that are comparable to those obtained using IL \citep{smith1969consolidation, sheahan1997experimental}. However, CRS completes the loading process substantially faster than IL \citep{wissa1971consolidation, lavorati2023evaluation}. We therefore adopt the CRS configuration in this work. 

\subsection{Staged Calibration}
\label{sec:stages}
 
This calibration procedure is inspired by our previous work on a multi-stage calibration method for continuum damage mechanics models \citep{saji2025multi}. This method identifies exponential damage law parameters sequentially by minimizing stage-specific loss functions defined using bespoke features of the force-displacement response. An earlier study applied a manual feature-based strategy \citep{mobasher2022modeling} to calibrate the shear-modified Gurson--Tvergaard--Needleman model \citep{nahshon2008modification}. In the present work, we adapt the concept of a multi-stage calibration procedure to the ECC model where the CS parameters are identified based on the responses of simulated triaxial and oedometer tests.

For each calibration sample, the simulated mechanical properties ($MS_{out}$) are compared with their reference values in the calibration dataset ($MS_{ref}$) through a normalized squared error, defined for the triaxial and oedometer tests as
\begin{equation}
O_{j} = \sum_{i \in j} \sqrt{\left( \frac{T_{i} - P_{i}}{T_{i}} \right)^{2}},
\qquad j \in \{\mathrm{tri}, \mathrm{oed}\},
\label{eq:calib_loss}
\end{equation}
where $T_{i}$ is the reference value of an mechanical property and $P_{i}$ is its simulated counterpart. $O_{\mathrm{tri}}$ represents loss function for $S_{u}$ and $E_{50}$ while $O_{\mathrm{oed}}$ represents the loss function for $p_{o}$ and $C_{c}$. The calibration proceeds one parameter at a time, where at stage $k$, a single parameter $x_{k}$ is varied within its admissible range $X_{k}$ and corresponding loss value is minimized,

\begin{equation}
x_{k} = \operatorname*{arg\,min}_{x_{k} \in X_{k}} O_{k},
\qquad k \in \{1, \dots, 5\},
\label{eq:calib_stage}
\end{equation}
with the five stages, their targets, and their bounds listed in
Table~\ref{tab:stages}. A stage concludes when its minimized loss satisfies
\begin{equation}
O_{k} \leq \theta_{k},
\label{eq:calib_pass}
\end{equation}
where $\theta_{k}$ is the stage-wise relative-error threshold that governs the transition to the next stage. $\theta_{k}$  is set to 20\% for every stage, and the cycle terminates when Eq. \eqref{eq:calib_pass} holds true for all $k \in \{1, \dots, 5\}$. The stage order is informed by a quantitative parametric sensitivity analysis of the effects of the CS parameters on the four mechanical properties, which is reported in the Supplementary material of this manuscript. The conclusions of this analysis are summarized in Table~\ref{tab:sensitivity}.
Accordingly, $M$ is calibrated first because it dominates $S_{u}$. Then, $\lambda$ and $\kappa$ are calibrated against $C_{c}$ and $E_{50}$ respectively. $M$ is refined next because updating $\kappa$ moderately shifts $S_{u}$. And finally, $a_{0}$ is calibrated last since it has no influence on the triaxial test and therefore does not vary the values of $S_{u}$ and $E_{50}$ obtained in earlier stages.

\begin{table}[htbp]
\caption{Order of the multi-stage calibration. Each stage varies one parameter within its bounds and minimizes the loss of the parameter it most directly controls.}
\label{tab:stages}
\centering
\small
\begin{tabular}{@{}clllc@{}}
\hline
Stage $k$ & Parameter varied & Simulated test & Bounds \\
\hline
1 & $M$  & triaxial & $0.01$ to $2.5$ \\
2 & $\lambda$ & oedometer & $10^{-4}$ to $1$ \\
3 & $\kappa$ & triaxial & $0.01$ to $0.5$ \\
4 & $M$ (refinement) & triaxial & $0.01$ to $2.5$ \\
5 & $a_{0}$ & oedometer & $10^{4}$ to $10^{6}$   \\
\hline
\end{tabular}
\end{table}

\begin{table}[htbp]
\caption{Summary of the conclusions from the quantitative parametric sensitivity analysis on the influence of each CS parameter on the four mechanical properties. The influence is classified as None, Low, Moderate, or High, and the governing parameter for each property is highlighted in bold. The complete quantitative results are provided in the Supplementary material. These conclusions informed the stage order adopted in this work (Table~\ref{tab:stages}).}
\label{tab:sensitivity}
\centering
\small
\begin{tabular}{@{}lcccc@{}}
\hline
 & $E_{50}$ & $S_{u}$ & $p_{o}$ & $C_{c}$ \\
\hline
$\kappa$  & \textbf{High}     & Moderate & None & None \\
$\lambda$ & None     & None     & Low  & \textbf{High} \\
$M$       & Moderate & \textbf{High}     & Low  & None \\
$a_{0}$   & None     & None     & \textbf{High} & Low  \\
\hline
\end{tabular}
\end{table}

The calibration procedure is written in Python, and the optimization algorithms are taken from the SciPy library   \citep{virtanen2020scipy}. Each stage is solved by the COBYLA algorithm  \citep{powell1994direct} from a fixed initial guess ($\kappa = 0.075, \lambda = 0.5, M = 1.0, a_{0} = 2 \times 10^{5}$ Pa), and every stage is capped at 25 forward evaluations to prevent runaway divergence. If a stage fails, it is retried with the Nelder-Mead method  \citep{nelder1965simplex} and then with a Brent bracketing search  \citep{brent1973algorithms}. In-stage stagnation triggers a restart from a perturbed starting point with an enlarged trust region. If non-convergence occurs despite this, the five stages are repeated for up to three cycles.  A cycle is defined as one pass through the five-stage sequence of Table~\ref{tab:stages}. For each sample, the calibration passes when all four mechanical properties lie within 20\% of their reference values, which is checked after every stage and at the end of every cycle. Samples that still fail are flagged and excluded from the envelope construction of Section~\ref{sec:envelope}. To ensure physical meaningfulness and numerical stability of the simulations, the conditions described in~\ref{apx:guards} are enforced. The complete
procedure for a single calibration dataset sample is summarized in Algorithm~\ref{alg:calibration}.

\begin{algorithm}[H]
\caption{Multi-stage calibration for a single calibration sample.}
\label{alg:calibration}
\KwIn{Mechanical properties reference set ($MS_{out} = (S_{u}, E_{50}, p_{o}, C_{c})$; 
basic properties $LL$, $PL$, $w$, $e$, $\gamma_{t}$; tolerance $\theta = 20\%$; cycle cap $N_{c} = 3$;}
\KwOut{Calibration set (CS) $(\kappa, \lambda, M, a_{0})$, or a
non-convergence flag}
Derive Fixed Set (FS)from the basic properties, the confining stress from $S_{u}$\;
Initialize CS and enforce the admissibility safeguards of Appendix~\ref{apx:guards}\;

  \For{$\mathrm{cycle} = 1, \dots, N_{c}$}{
    \For{$\mathrm{k} = 1, \dots, 5$ (Table~\ref{tab:stages})}{
      $x_{k} \gets \operatorname*{arg\,min}_{x_{k} \in X_{k}} O_{k}$ \;
      \eIf{$O_k < \theta_k$}{
        \Return converged CS parameter\;
        }{
       \Return stage non-convergence flag\;
      }
    }
    \eIf{$O_{k} \leq \theta_{k} \;\; \forall\, k \in \{1, \dots, 5\}$}{
      \Return converged CS\;
    }{
    Update CS bounds \;
    }
  }
\Return Sample non-convergence flag \tcp*{sample excluded from the predictive envelope}
\end{algorithm}
 
\section{Predictive Envelope}
\label{sec:envelope}
 
\begin{figure}[H]
\centering
\includegraphics[width=1\textwidth,trim = 0cm 1.75cm 0cm 3cm, clip]{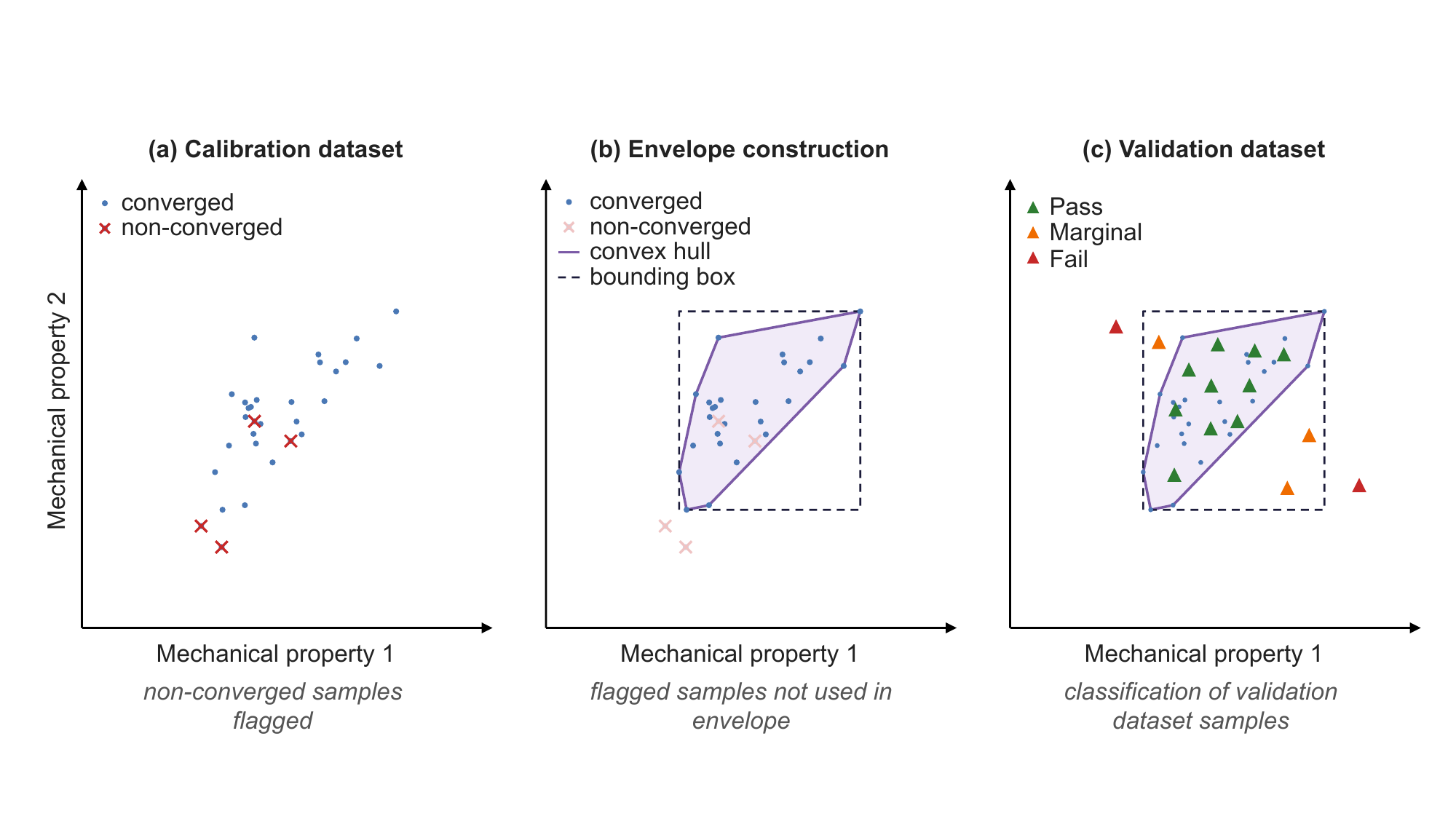}
\caption{Construction and use of the predictive envelope, shown schematically with synthetic data. (a) The calibration dataset, where non-converged samples are marked and excluded. (b) The bounding box and convex hull constructed from the converged calibration samples alone, with the excluded samples shown
faded. (c) Validation samples superimposed and classified as pass when they fall inside the convex hull, marginal when they fall inside the bounding box but not convex hull, and fail when they fall outside both.}
\label{fig:envelope_construction}
\end{figure}

This section describes the predictive envelope against which the validation dataset is assessed. The envelope is constructed for every Cell produced by the binning procedure described in Section~\ref{sec:binning}. 
The envelope construction begins by first flagging the non-converged samples in the calibration dataset as shown in Fig.~\ref{fig:envelope_construction}a. These are then excluded from the rest of the procedure.
 
Then, as shown in Fig.~\ref{fig:envelope_construction}b, the converged
samples are plotted as points in the mechanical property space of each
test, namely the plane of $S_{u}$ and $E_{50}$ for the triaxial test and of
$p_{o}$ and $C_{c}$ for the oedometer test. Two nested regions are drawn
around the points. The bounding box is the rectangle that spans the extrema
of the two properties. The convex hull is the smallest convex polygon that
encloses all points and is computed with the Quickhull algorithm
\citep{barber1996quickhull}. By construction, the convex hull lies within
the bounding box, and thus serves as a stricter assessment criterion.
 
Finally, validation samples are plotted as points in the same space and classified as pass when its point falls inside the convex hull, marginal when it falls inside the bounding box only, and fail when it falls outside both. Fig.~\ref{fig:envelope_construction}c illustrates these classifications visually. The hull and the bounding box thereby provide the basis for assessing how well the calibration dataset represents the remainder of the Cell. We quantify this representativeness through an accuracy variable, which captures the coverage of the calibrated envelope for a Cell, defined as
\begin{equation}
A_j = 
\frac{\max_{i \in S} x_{ij} \;-\; \min_{i \in S} x_{ij}}
     {\max_{i \in \mathcal{C}} x_{ij} \;-\; \min_{i \in \mathcal{C}} x_{ij}}
\ \times 100,
\qquad
\bar{A} = \frac{1}{m}\sum_{j=1}^{m} A_j ,
\label{eq:accuracy}
\end{equation}
where $A_{j}$ and $\bar{A}$ denote the accuracy of mechanical property $j$ and the mean accuracy over the $m$ mechanical properties considered, respectively, $x_{ij}$ denotes the magnitude of property $j$ for sample $i$, and $S$ denotes the converged samples of the calibration dataset drawn from Cell $\mathcal{C}$. The accuracy $A_{j}$ thus measures the fraction of the Cell range of property $j$ that the converged calibration samples cover. A higher value of $A_{j}$ indicates that the calibration dataset is more representative of the Cell.

%%%%%%%%%%%%%%%%%%%%%%%%%%%%%%%%%%%%%%%%%%%%%%%%%%%
\section{Demonstration of the BRIDGE workflow}
\label{sec:demo}

This section demonstrates the complete BRIDGE workflow for a representative
cell and presents the intermediate outputs in the order in which they are
generated. These comprise of (a) the automated multi-stage calibration of the
selected samples, (b) the construction of the predictive envelope from the
successfully calibrated samples, (c) the mapping of the validation samples
onto this envelope, and (d) the prediction of mechanical properties of new samples. 
Unless stated otherwise, this demonstration and subsequent numerical studies
(Section~\ref{sec:numerical}) use the screened dataset of 2,609 samples and the
\textbf{EC} partitioning strategy described in Section~\ref{sec:workflow}.
The forward models use a single axisymmetric CAX8RP element as described in Section~\ref{sec:forward}. The stage-wise and global tolerances are set to \(\theta=20\%\), and the calibration is limited to \(N_c=3\) cycles. All computations are performed on a Dell Precision 7920 Tower with a 2.1~GHz Intel\textsuperscript{\textregistered}
Xeon\textsuperscript{\textregistered} Gold 6230R processor and 64~GB of RAM,
rated at 873.6~GFLOPS \citep{intel_processors}.

\begin{figure}[H]
\centering
\includegraphics[width=1\textwidth,trim = 0cm 0cm 0cm 0cm, clip]{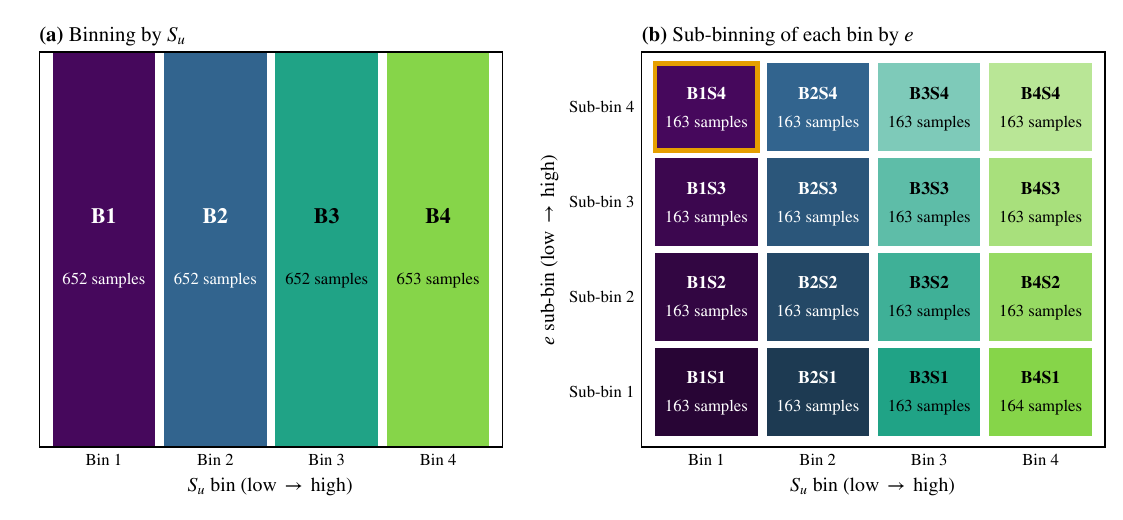}
\caption{Partition of the screened dataset into Cells. (a) The 2,609 samples are first divided into four \textbf{EC} bins by the undrained shear strength $S_{u}$. (b) Each Bin is then subdivided into four \textbf{EC} sub-bins by the void ratio $e$, which yields 16 Cells labeled B$i$S$j$, where $i$ denotes the $S_{u}$ Bin and $j$ denotes the $e$ Sub-Bin. Each Cell reports its number of samples, the color families distinguish the $S_{u}$ bins. The orange outline marks Cell B1S4, the fourth $e$ Sub-Bin of the first $S_{u}$ Bin, which serves as the representative Cell used for demonstration.}
\label{fig:Demo_dataset_partition}
\end{figure}

For this demonstration, the dataset is divided into four \(S_u\) bins, as shown in
Fig.~\ref{fig:Demo_dataset_partition}a. These bins are further subdivided into four \(e\) sub-bins, as shown in Fig.~\ref{fig:Demo_dataset_partition}b. This produces 16 cells
containing 163 or 164 samples each. Cell~B1S4, is selected as the representative cell.
Of its 163 samples, \(n_s=30\) (\(18.4\%\)) are selected for calibration using
Puchwein sampling, and the remaining 133 samples form the validation dataset. The same cell is also used in Study~C (Section \ref{sec:studyC}).

\subsection{Automated multi-stage calibration}
\label{sec:demo_calibration}
 
\begin{figure}[H]
  \centering
 \includegraphics[width=\textwidth]{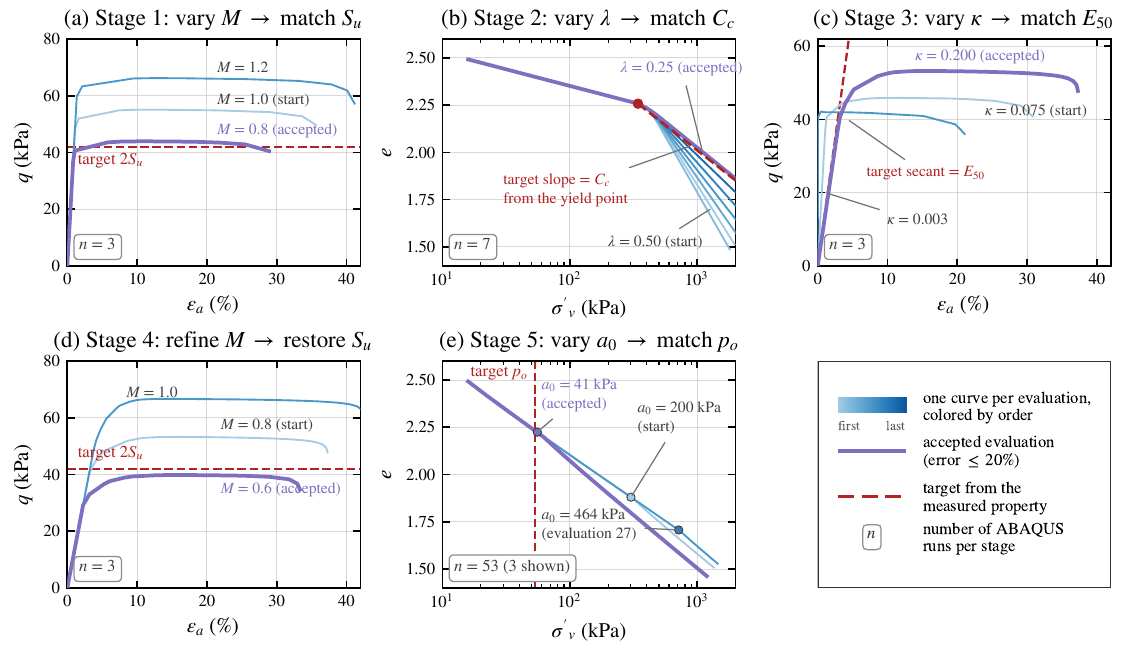}
  \caption{Simulated responses over the five calibration stages of Table~\ref{tab:stages} for a representative calibration sample of Cell B1S4. One stage is shown per panel from (a) to (e), where each curve represents one finite element evaluation and is labeled
  with the parameter value used therein. Successive evaluations are colored by order from light to dark, the accepted evaluation of each stage is drawn in bold purple, the red dashed line marks the target of the calibration stage obtained from the reference property of the sample. Each panel reports the number of finite element evaluations $n$ used in the stage.
  (a) Stage 1 varies $M$ until the deviatoric stress plateau of the triaxial response
  matches $2S_u$. (b) Stage 2 varies $\lambda$ until the virgin compression slope of the
  CRS oedometer curve coincides with the target slope $C_c$ anchored at that point. (c) Stage 3 varies $\kappa$ to match $E_{50}$. (d) Stage 4 refines $M$ to restore the $S_u$ match perturbed by stage 3. (e) Stage 5 varies $a_0$ until it matches $p_o$. Only selected finite element responses are displayed for clarity.}
  \label{fig:demo_calib_stages}
\end{figure}

Fig.~\ref{fig:demo_calib_stages} presents the five calibration stages of
Table~\ref{tab:stages} for one calibration sample of the Cell B1S4. Each panel corresponds to
one stage. Each curve represents one finite element evaluation. Stage 1 varies $M$ until
the deviatoric stress plateau of the simulated triaxial response matches the target
$2S_u$, as shown in Fig.~\ref{fig:demo_calib_stages}a. Stage 2 varies $\lambda$ until
the virgin compression slope of the oedometer curve matches the measured
$C_c$, as shown in Fig.~\ref{fig:demo_calib_stages}b. Stage 3 varies $\kappa$ until the
secant stiffness matches $E_{50}$, as shown in Fig.~\ref{fig:demo_calib_stages}c. This
adjustment affects the $S_u$ value obtained in Stage 1. Stage 4 therefore refines $M$ and restores the $S_u$ match, as shown in Fig.~\ref{fig:demo_calib_stages}d. Finally, stage 5 varies $a_0$ until it matches the reference $p_o$ value, as shown in Fig.~\ref{fig:demo_calib_stages}e. 

\begin{figure}[H]
  \centering
   \includegraphics[width=\textwidth]{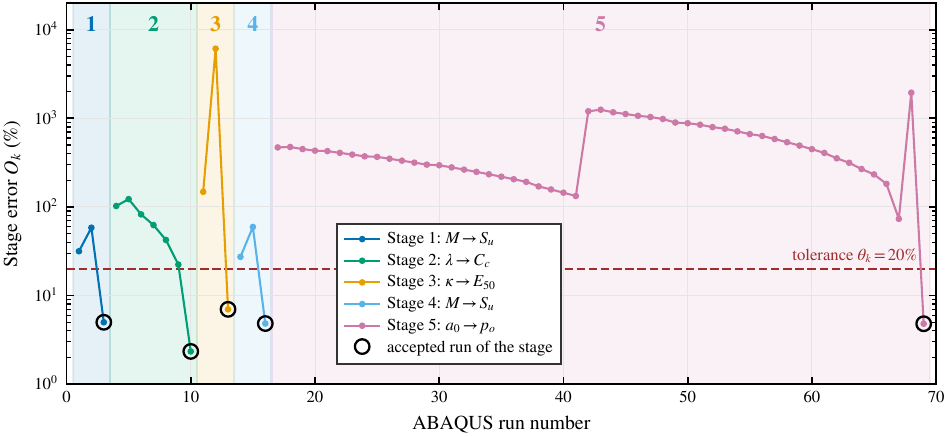}
\caption{Stagewise error history of the first calibration sample of Cell B1S4. Each marker denotes one ABAQUS run, colored by the calibration stage indicated by the shaded bands. The open circles mark the accepted ABAQUS run, i.e., the first run with a stage-wise error $O_k$ below the dashed line which marks the convergence tolerance $\theta_{k}=20\%$.}
  \label{fig:demo_calib_convergence}
\end{figure}

Fig.~\ref{fig:demo_calib_convergence} reports the relative error of each stage target
against the ABAQUS run number. Every
stage continues until its error is below the acceptance threshold $\theta_k = 20\%$. No manual intervention occurs at any point during the calibration process. Applied to the full calibration dataset of the Cell, 27 of the 30 selected samples
converge. The remaining three samples terminate with a non-convergence flag, with residual mismatches on $E_{50}$ (41.7\%), $p_o$ (50.7\%), and
$C_c$ (21.2\%), respectively. The flags identify the failing property and stage of each
sample, which supports the failure analysis of Section~\ref{sec:reliability}.
 
\subsection{From calibrated samples to the predictive envelope}
\label{sec:demo_envelope}
 
\begin{figure}[H]
  \centering
   \includegraphics[width=\textwidth]{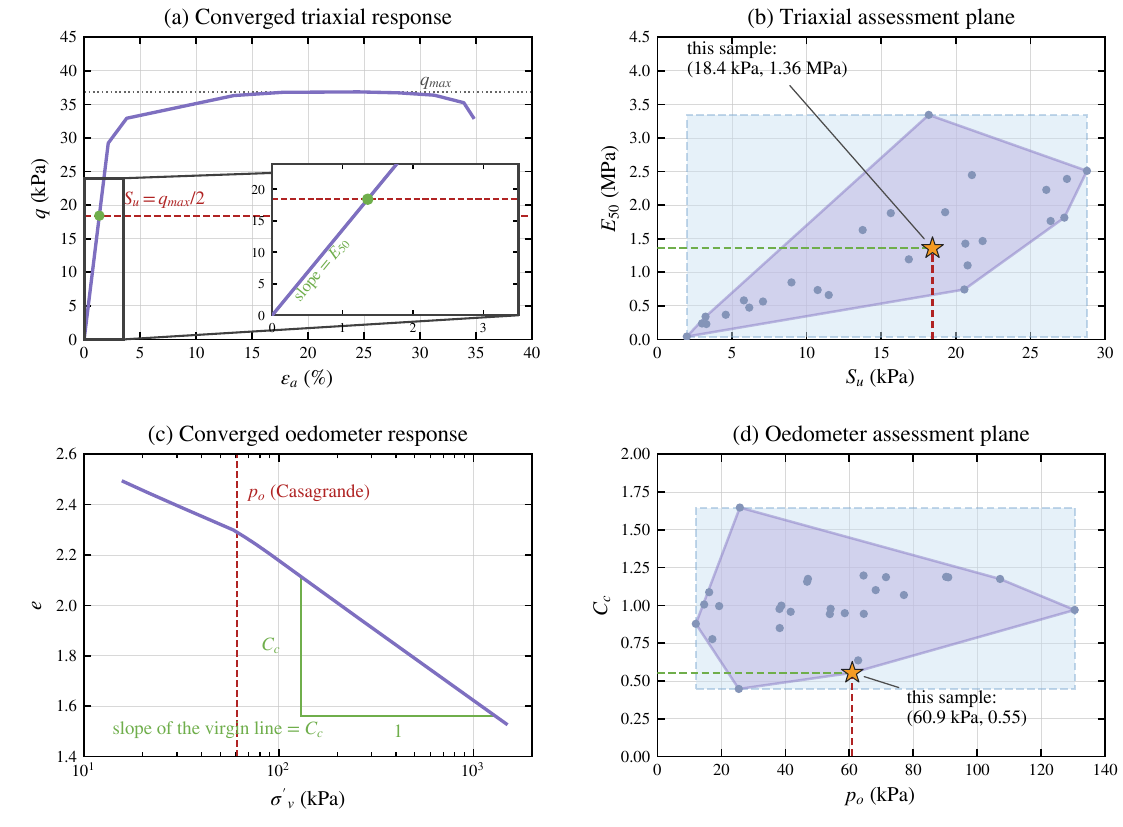}
  \caption{Contribution of a successfully calibrated sample to the predictive envelope of Cell B1S4. Panels (a) and (c) show the converged triaxial and oedometer responses, respectively, for the representative sample in Fig.~\ref{fig:demo_calib_stages}. Panel (a) shows the extraction of $S_u$ and $E_{50}$. The inset magnifies the initial response and shows the secant slope used to determine $E_{50}$. Panel (c) shows the extraction of $p_o$ using the Casagrande method and $C_c$ from the slope of the virgin compression line. Panels (b) and (d) map the resulting ($S_u$, $E_{50}$) and ($p_o$, $C_c$) pairs, respectively, onto the corresponding mechanical property spaces. The star represents this sample, while the other markers represent the remaining successfully calibrated samples in the Cell. Together, the successfully calibrated samples define the convex hull and bounding box for cel B1S4.}
  \label{fig:demo_sample_to_point}
\end{figure}

\begin{figure}[H]
  \centering
   \includegraphics[width=\textwidth]{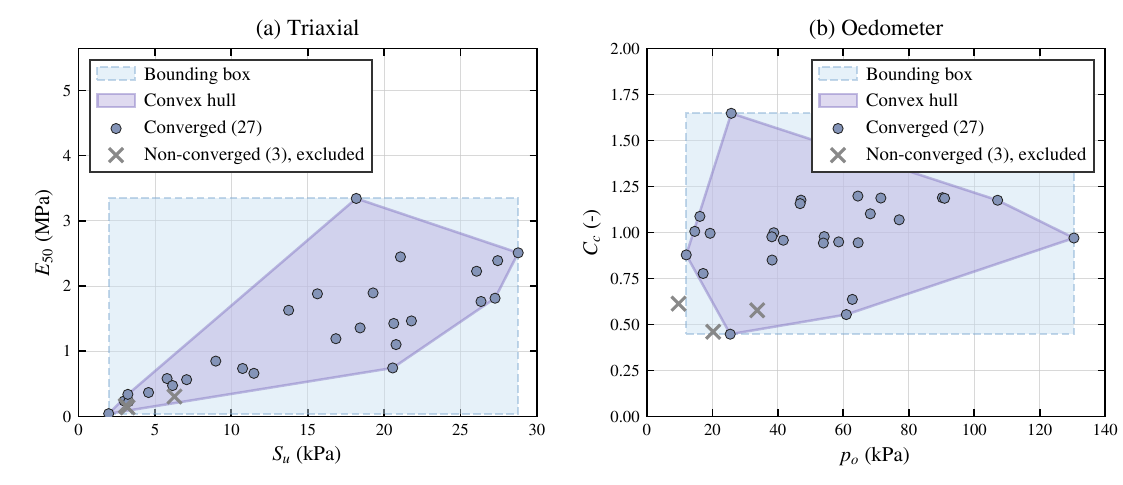}
  \caption{Construction of the predictive envelope of Cell B1S4 in the triaxial (left) and oedometer (right) assessment planes. The 27
  converged calibration samples are shown as filled markers and the three non-converged
  samples as faded crosses, and the bounding box (dashed) and convex hull (solid) are
  constructed from the converged samples alone, whereby the flagged samples are excluded
  from the envelope.}
  \label{fig:demo_envelope_construction}
\end{figure}

Fig.~\ref{fig:demo_sample_to_point} shows how a calibrated sample contributes one point
to each assessment plane of the predictive envelope. The calibration terminates when the error $O_k$ falls below the $20\%$ threshold for all stages. From the triaxial response, $S_u$ is extracted as half the peak deviatoric
stress and $E_{50}$ as the secant slope at half that peak, as shown in
Fig.~\ref{fig:demo_sample_to_point}a. From the oedometer response, $p_o$ is extracted by
the Casagrande method and $C_c$ as the slope of the virgin compression line, as shown in
Fig.~\ref{fig:demo_sample_to_point}c. The pair ($S_u$, $E_{50}$) defines the coordinates
of the sample in the triaxial assessment plane, as shown in
Fig.~\ref{fig:demo_sample_to_point}b. The pair ($p_o$, $C_c$) defines its coordinates in
the oedometer assessment plane, as shown in Fig.~\ref{fig:demo_sample_to_point}d. Fig.~\ref{fig:demo_envelope_construction} presents the final predictive envelope Cell B1S4. The 27 converged samples populate the triaxial and oedometer assessment planes. The three non-converged samples appear as faded crosses. They are flagged and excluded from the envelope construction. The bounding box and the convex hull of Section~\ref{sec:envelope} are constructed from the converged points alone. 
 
\subsection{Validation mapping and the calibrated output}
\label{sec:demo_validation}
 
\begin{figure}[H]
  \centering
  \includegraphics[width=\textwidth]{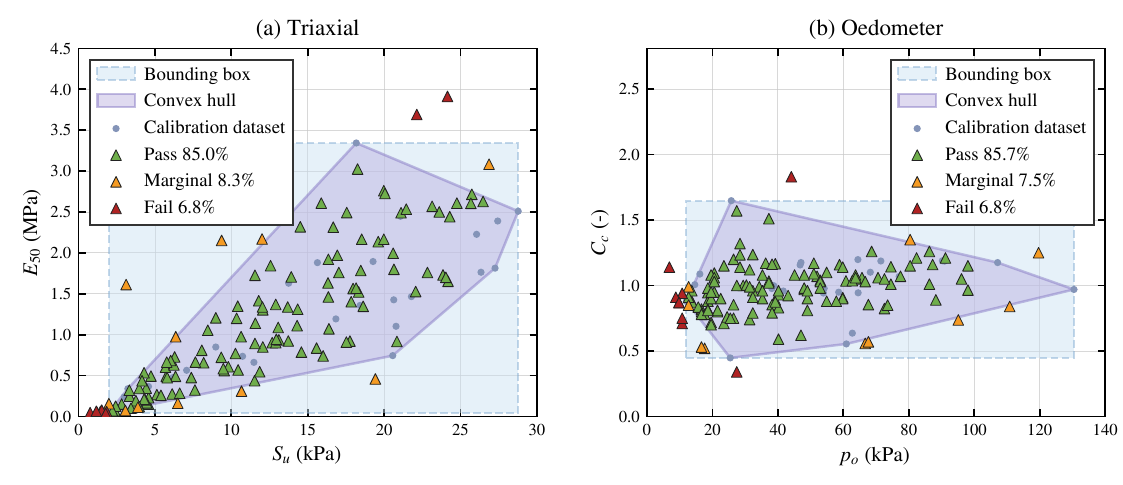}
  \caption{Validation of the predictive envelope of Cell B1S4, in which the measured
  mechanical properties of the 133 validation samples are cross-plotted onto the triaxial
  (left) and oedometer (right) assessment planes and classified as pass, marginal, or fail
  with respect to the convex hull and bounding box constructed in
  Fig.~\ref{fig:demo_envelope_construction}.}
  \label{fig:demo_validation}
\end{figure}

Fig.~\ref{fig:demo_validation} presents the validation of the predictive envelope. The measured mechanical properties of the 133 validation dataset samples are cross-plotted onto the triaxial and oedoemter assessment
planes. A sample passes when it falls inside the convex hull. It is classified as marginal when it falls inside the bounding box but not the convex hull. It
fails when it falls outside the bounding box. The resulting pass, marginal, and fail fractions are 85.0, 8.3, and 6.8\% in the triaxial plane and 85.7, 7.5, and 6.8\% in the oedometer plane, respectively. If a majority of the validation samples are classified as pass or marginal, as is the case here, the Cell is considered validated. The CS parameter ranges of the Cell are obtained from the converged calibration samples and are listed in Table~\ref{tab:parameter_range}. This result demonstrates that using the BRIDGE framework, a small calibration dataset can sufficiently represent the entire Cell.
 
\begin{table}[H]
\caption{Calibrated CS parameter ranges of the representative Cell B1S4, obtained from the converged calibration samples of the Cell.}
\label{tab:parameter_range}
\centering
\begin{tabular}{@{}lcl@{}}
\hline
Parameter & Calibrated range\\
\hline
$\kappa$  & 0.049 to 0.42\\
$\lambda$ & 0.20 to 0.75\\
$M$       & 0.60 to 0.80\\
$a_{0}$   & 7 to 143 kPa\\
\hline
\end{tabular}
\end{table}

\subsection{Prediction}

Once all the Cells of the dataset have been validated, they can be used to
predict the mechanical properties of a new sample for which only the basic 
properties and $S_u$ are available. The prediction proceeds in two steps:
\begin{enumerate}
    \item \textbf{Cell assignment:} the measured void ratio $e$ and undrained
    shear strength $S_u$ of the new sample are compared with the bin and sub-bin
    boundaries of the dataset to assign the sample to the corresponding validated
    Cell.
    \item \textbf{Property prediction:} the mechanical properties of the new
    sample are predicted as a range. This range is that spanned by the mechanical properties ($MS_{out}$) of the converged calibration samples of the assigned Cell.
\end{enumerate}

\begin{figure}[H]
  \centering
  \includegraphics[width=\textwidth]{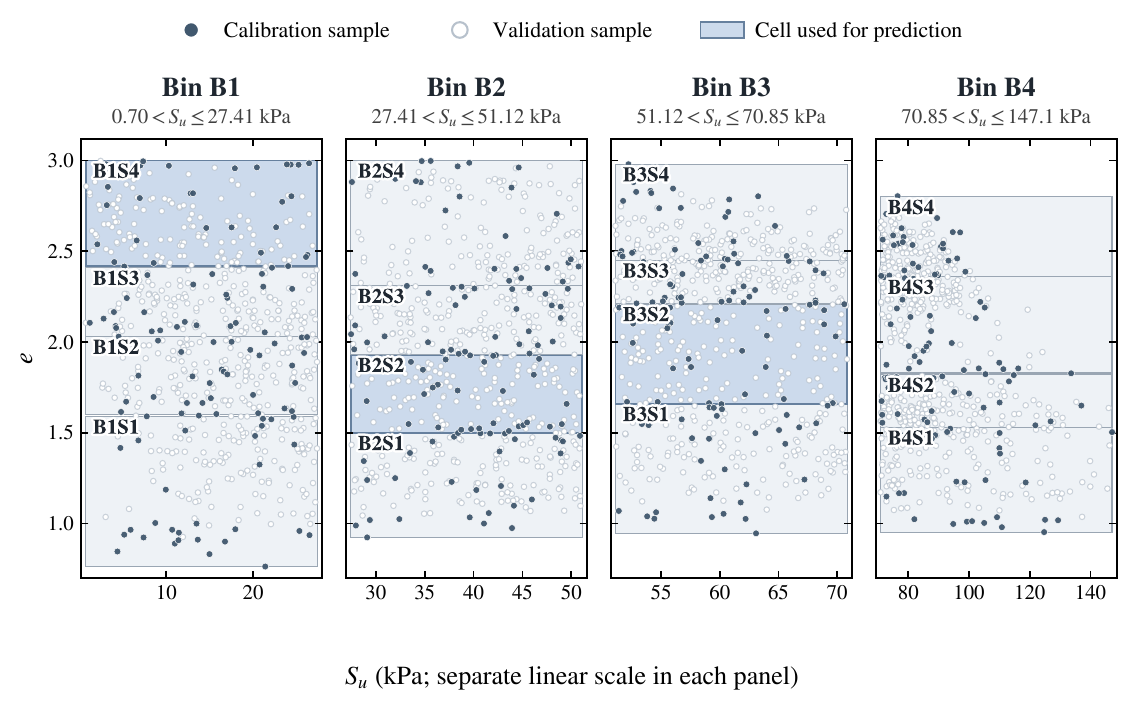}
  \caption{The sixteen validated Cells considered in this demonstration study are represented in the \((S_u,e)\) plane, where each \(S_u\) bin is displayed on an independent linear scale for visual clarity. Filled and open circles denote calibration and validation samples, respectively. The shaded Cells B1S4, B2S2, and B3S2 are the Cells used to demonstrate the prediction process.}
  \label{fig:cellgrid}
\end{figure}

%========================================================================
Fig.~\ref{fig:cellgrid} shows the sixteen validated Cells, B1S1 to B4S4, considered here, which are arranged as four $S_u$ bins, each subdivided into four $e$ sub-bins according to the \textbf{EC} scheme. We demonstrate the prediction on three Cells, namely B1S4, B2S2, and B3S2, where the 133 validation samples of each Cell serve as the prediction samples. Only the $S_u$ and $e$ of the prediction samples are used to assign them to these Cells, while their measured $E_{50}$, $p_o$, and $C_c$ are used solely to assess the prediction.

\begin{figure}[H]
    \centering
    \includegraphics[width=\textwidth]{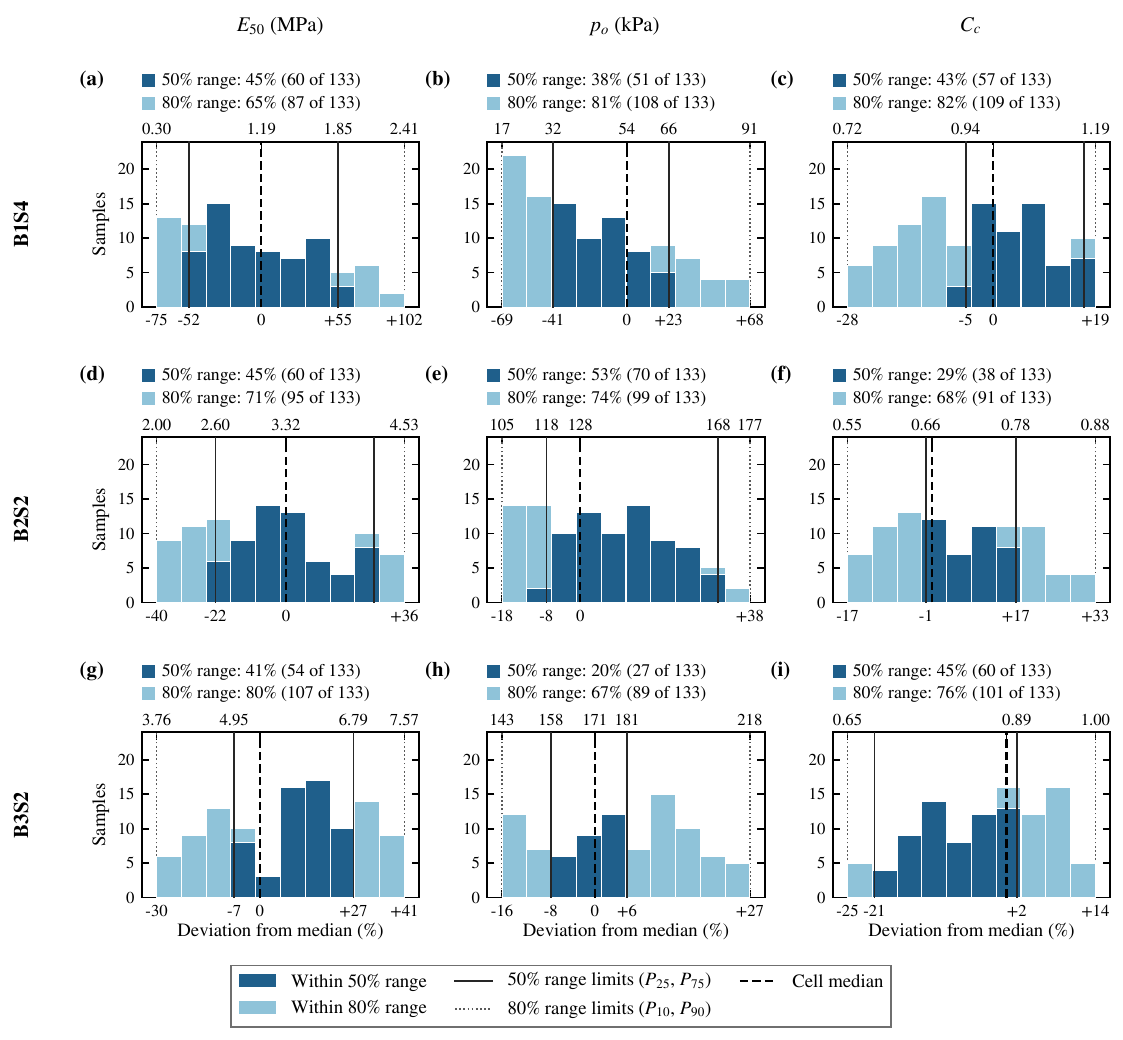}
    \caption{Prediction samples of Cells B1S4 (a to c), B2S2 (d to f), and B3S2 (g to i) that fall within the predicted ranges of $E_{50}$, $p_o$, and $C_c$ about the Cell median. Each panel spans the 80\% range, bounded by $P_{10}$ and $P_{90}$ of the calibrated values of the converged calibration samples, where the light bars represent all prediction samples within this range and the darker bars represent those within the 50\% range, bounded by $P_{25}$ and $P_{75}$. The lower axis gives the deviation from the Cell median, the upper axis gives the corresponding values in physical units, and the vertical axis gives the number of prediction samples in each bin. The panel titles report the share of the 133 prediction samples of the Cell contained in each range.}
    \label{fig:prediction_median}
\end{figure}

\begin{figure}[H]
    \centering
    \includegraphics[width=\textwidth]{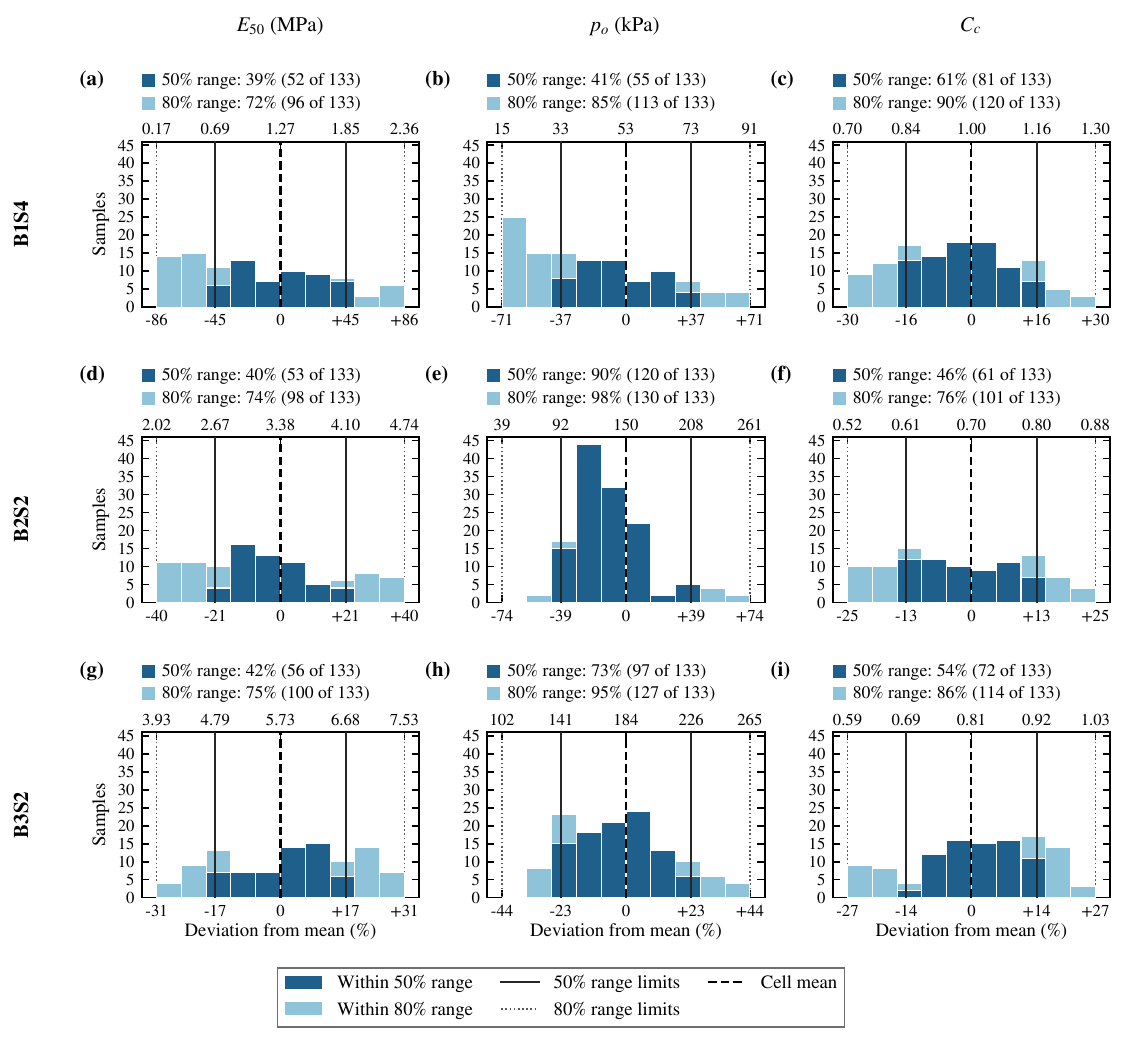}
    \caption{Prediction samples of Cells B1S4 (a to c), B2S2 (d to f), and B3S2 (g to i) that fall within the predicted ranges of $E_{50}$, $p_o$, and $C_c$ about the Cell mean. Each panel spans the 80\% range, bounded by the mean of the calibrated values of the converged calibration samples $\pm\,1.28\sigma$, where the light bars represent all prediction samples within this range and the darker bars represent those within the 50\% range, bounded by the mean $\pm\,0.67\sigma$. Here $\sigma$ is the standard deviation of the calibrated values, and both ranges assume a normal, bell-shaped distribution. The lower axis gives the deviation from the Cell mean, the upper axiss gives the corresponding values in physical units, and the vertical axis gives the number of prediction samples in each bin. The panel titles report the share of the 133 prediction samples of the Cell contained in each range.}
    \label{fig:prediction_mean}
\end{figure}

For each Cell and mechanical property, we construct the 50\% and 80\% prediction ranges from the calibrated values of the converged calibration samples in two ways. The first uses percentiles about the median, where the 50\% range spans $P_{25}$ to $P_{75}$ and the 80\% range spans $P_{10}$ to $P_{90}$. The second uses the mean and the standard deviation $\sigma$ of the calibrated values, where the 50\% and 80\% ranges are the mean $\pm\,0.67\sigma$ and $\pm\,1.28\sigma$, respectively. This second construction adopts the general assumption of a normal, bell-shaped distribution, for which these ranges enclose 50\% and 80\% of the values. Figs.~\ref{fig:prediction_median} and~\ref{fig:prediction_mean} present the prediction samples contained in each range, and Table~\ref{tab:coverage} summarizes the corresponding predictions. The prediction ranges satisfy the evaluation criterion for the predictions stated in Section~\ref{sec:overview}. On average over all the Cell and property combinations, the 50\% ranges contain 40 and 54\% of the prediction samples for the median and mean based constructions, respectively while the 80\% ranges contain 74 and 83\% of the prediction samples, respectively (Table~\ref{tab:coverage}). For the 50\% range, fourteen of the eighteen cases contain between 38
and 61\% of the prediction samples, while of the remaining four cases only two fall below 30\%. For the 80\% range, all eighteen cases contain at least 65\% of the prediction samples. The BRIDGE framework therefore delivers consistent predictions across three Cells, three mechanical properties, and two prediction range constructions, thereby satisfying the consistency objective stated in Section~\ref{sec:overview}.

It is worth noting that the predicted ranges remain wide in physical terms, particularly for $E_{50}$ and $p_o$. In B1S4, for example, the 80\% range about the median spans 0.30 to 2.41~MPa for $E_{50}$ and 17 to 91~kPa for $p_o$. This width is inherited from the dataset itself, as two observations confirm. First, Fig.~\ref{fig:cellgrid} shows that the samples within each Cell are widely distributed, with $S_u$ in B1S4 varying from 0.76 to 26.9~kPa. Second, Figs.~\ref{fig_parameter_relations} and~\ref{fig_Parameter_heatmap_spearman} show that $E_{50}$ and $p_o$ correlate only weakly with the basic properties, with $|\rho|$ not exceeding 0.38. Reducing the width of the predicted mechanical property ranges, which stems from the inherent scatter of the dataset, therefore remains an avenue for future work.

\begin{table}[H]
    \centering
    \caption{Percentage of the 133 validation samples of each Cell that fall within the
    predicted 50\% and 80\% ranges. The median based ranges are the central percentile
    ranges $P_{25}$ to $P_{75}$ and $P_{10}$ to $P_{90}$ of the calibrated values of the
    converged calibration samples, and the mean based ranges are the mean of those
    values $\pm\,0.67\sigma$ and $\pm\,1.28\sigma$, respectively.}
    \label{tab:coverage}
    \begin{tabular}{llcccc}
        \toprule
        & & \multicolumn{2}{c}{50\% range} & \multicolumn{2}{c}{80\% range} \\
        \cmidrule(lr){3-4}\cmidrule(lr){5-6}
        Cell & Property & Median & Mean & Median & Mean \\
        \midrule
        B1S4 & $E_{50}$  & 45 & 39 & 65 & 72 \\
             & $p_o$     & 38 & 41 & 81 & 85 \\
             & $C_c$     & 43 & 61 & 82 & 90 \\
        \midrule
        B2S2 & $E_{50}$  & 45 & 40 & 71 & 74 \\
             & $p_o$     & 53 & 90 & 74 & 98 \\
             & $C_c$     & 29 & 46 & 68 & 76 \\
        \midrule
        B3S2 & $E_{50}$  & 41 & 42 & 80 & 75 \\
             & $p_o$     & 20 & 73 & 67 & 95 \\
             & $C_c$     & 45 & 54 & 76 & 86 \\
        \midrule
        \multicolumn{2}{c}{\textbf{Average}} & \textbf{40} & \textbf{54} & \textbf{74} & \textbf{83} \\
        \bottomrule
    \end{tabular}
\end{table}
%=========================================================================

\begin{figure}[H]
\centering
\includegraphics[width=\textwidth, trim = 0cm 0cm 0cm 0cm, clip]{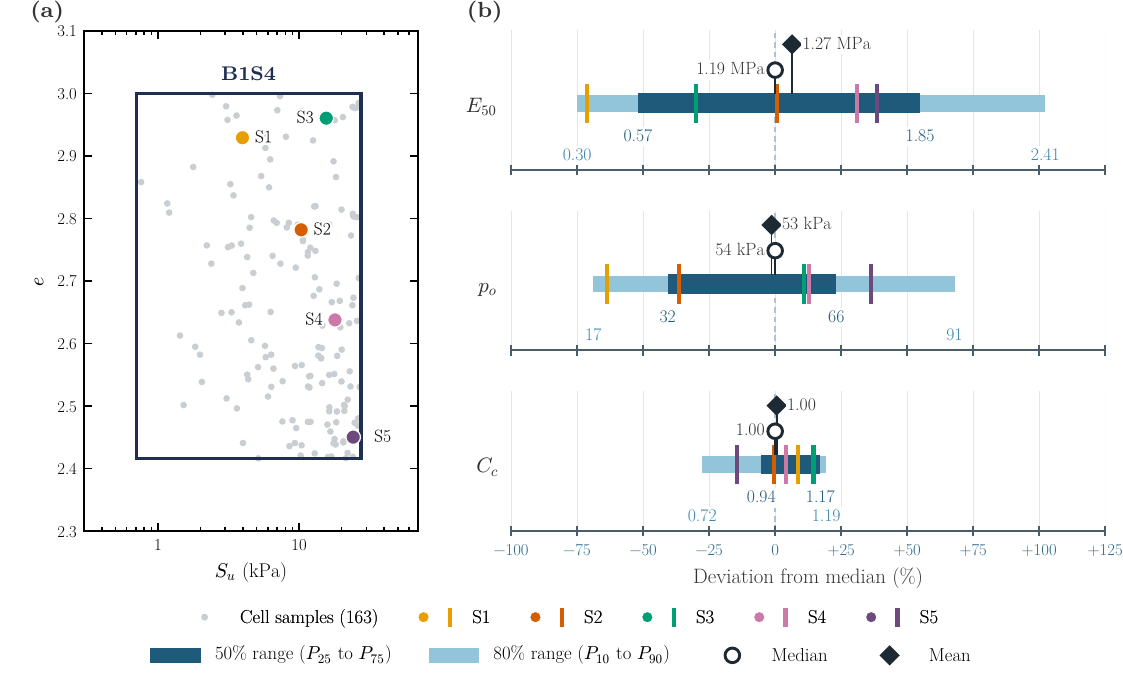}
\caption{Illustrative prediction for Cell B1S4. (a) The 163 samples of the Cell, in the $(S_u, e)$ plane, where the box marks the Cell boundaries and S1 to S5 denote five randomly selected prediction samples. (b) Predicted ranges of $E_{50}$, $p_o$, and $C_c$ constructed from the calibrated values of the 27 converged calibration samples and expressed as the deviation from the median, where the dark and light bars denote the 50\% ($P_{25}$ to $P_{75}$) and 80\% ($P_{10}$ to $P_{90}$) ranges, respectively. The circle and diamond symbols represent the median and mean values of the mechanical property. The colored ticks mark the measured mechanical properties of prediction samples S1 to S5, which allows a direct comparison of where each measured value lies with respect to the predicted 50\% and 80\% ranges.}
\label{fig:B1S4_prediction}
\end{figure}

Finally, to demonstrate the practical use of the predicted ranges, Fig.~\ref{fig:B1S4_prediction} compares the prediction ranges of Cell B1S4 with the measured mechanical properties of five prediction samples, S1 to S5, which were selected to span the $e$ range of the Cell. Fig.~\ref{fig:B1S4_prediction}(a) visualizes the five samples among the 163 samples of the Cell in the in the $(S_u, e)$ plane, while Fig.~\ref{fig:B1S4_prediction}(b) compares their measured mechanical properties with the 50\% and 80\% ranges about the median, where the mean is also marked marked for reference. The measured values of all five samples lie within the 80\% range for all three properties, while those of S2, S3, and S4 also lie within the 50\% range. Thus, the predicted ranges of a Cell provide a reasonable estimate of the mechanical properties of a new sample from its $S_u$ and basic properties alone, and this holds true across the $e$ span of the Cell.

The stages demonstrated here constitute one complete pass of the BRIDGE workflow. Studies A to C in Section \ref{sec:numerical} quantify how its accuracy and computational cost are affected by the binning criterion, the binning level, and the sampling design.

%%%%%%%%%%%%%%%%%%%%%%%%%%%%%%%%%%%%%%%%%%%%%%%%%%%%%%%%%%%%%%%%%%%%%%%%%%%%%%%%%%%%

%%%%%%%%%%%%%%%%%%%%%%%%%%%%%%%%%%%%%%%%%%%%%%%%%%%%%
\section{Numerical studies}
\label{sec:numerical}

This section examines different aspects of the
BRIDGE framework through three numerical studies, whose designs are summarized
in Table~\ref{tab:studies}. Study~A (Section~\ref{sec:studyA}) evaluates the
choice of binning criterion. Study~B (Section~\ref{sec:studyB}) examines the
effect of introducing a second binning level (sub-binning), while
Study~C (Section~\ref{sec:studyC}) investigates the influence of the sampling
method and sample budget. For each study, the adopted binning
configuration and accuracy values are reported. Unless stated otherwise, all analyses use the dataset, forward model, tolerances $\theta$, cycle cap $N_c$, and sampling method specified in Section~\ref{sec:demo}. Finally, Section~\ref{sec:reliability} collates the findings of the three studies to assess the reliability of the calibration procedure and the overall predictive accuracy of the BRIDGE framework.

\begin{table}[htbp]
\caption{Design of the three numerical studies. The sampling method and sample budget are held fixed unless the study varies them.}
\label{tab:studies}
\centering
\footnotesize
\begin{tabular}{@{}llp{4.6cm}c@{}}
\hline
Study & Aspect demonstrated & Design & Calibrations \\
\hline
A & binning criterion & 8 quantile bins on $S_{u}$ and on $e$ & 16 \\
B & sub-binning & $4 \times 4$ binning and sub-binning on $S_{u}$ and $e$ & 16 \\
C & sampling method and sample budget & five methods (Section~\ref{sec:sampling}), $n_{s} \in \{20, 30, 40\}$, one representative level-two Cell & 15 \\
\hline
\end{tabular}
\end{table}

\subsection{Study A: Binning Criterion}
\label{sec:studyA}
\begin{figure}[H]
  \centering
  \includegraphics[width=\textwidth]{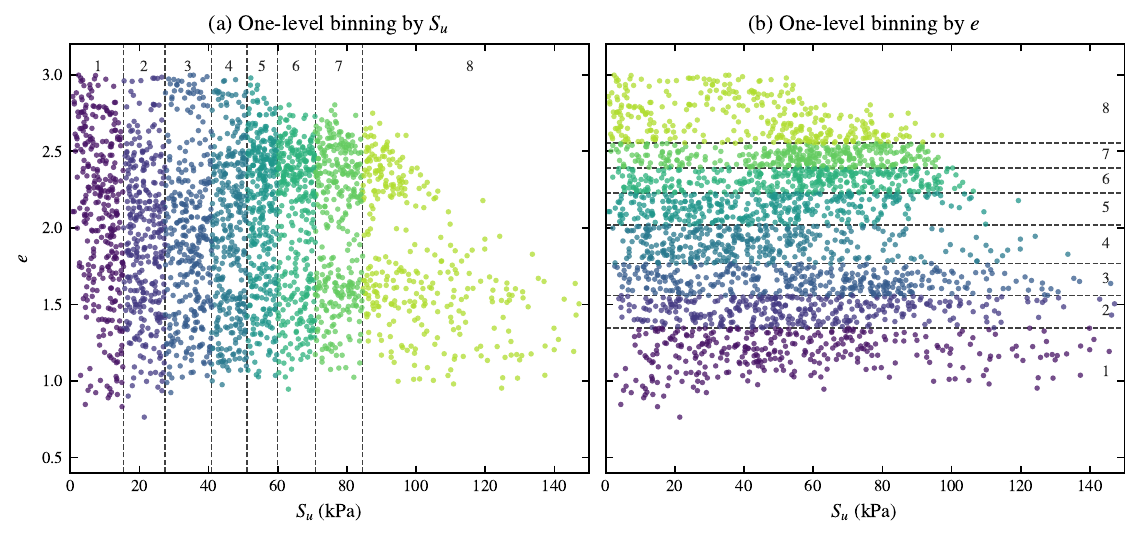}
  \caption{Comparison of one-level \textbf{EC} binning using $S_u$ and $e$. The same samples are shown in the $S_u$-$e$ property space. (a) Binning by $S_u$, with vertical dashed lines marking the Bin boundaries. (b) Binning by $e$, with horizontal dashed lines marking the Bin boundaries. Each criterion partitions the working dataset into 8 bins containing approximately equal numbers of samples. Colors and numerical labels indicate the corresponding Bin assignments.}
  \label{fig:Study_A_binning}
\end{figure}

Here, we examine the choice of the binning criterion. Two binning criteria are used in this study a) undrained shear strength $S_{u}$ and b) void ratio $e$ to divide the dataset into eight \textbf{EC} bins as shown in Fig.~\ref{fig:Study_A_binning}. The study therefore involves the calibration of 16 one-level bins. Of the 480 calibration samples used in this study, 464 (96.7\%) converge.

\begin{figure}[H]
\centering
\begin{subfigure}[t]{0.51\textwidth}
  \centering
  \includegraphics[width=\textwidth]{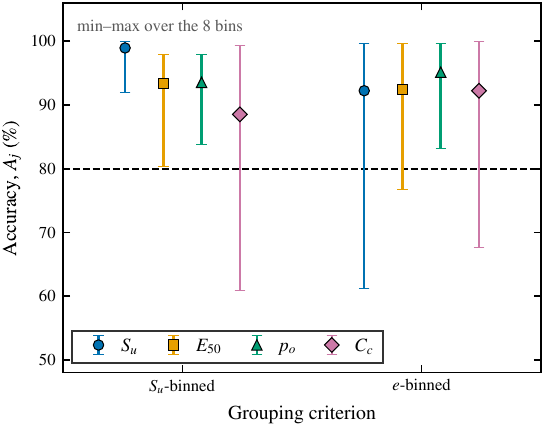}
  \caption{}
  \label{fig:studyA_coverage}
\end{subfigure}
\hfill
\begin{subfigure}[t]{0.46\textwidth}
  \centering
  \includegraphics[width=\textwidth]{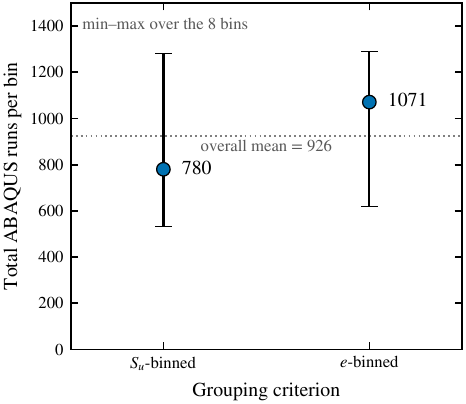}
  \caption{}
  \label{fig:studyA_cost}
\end{subfigure}
\caption{Accuracy and computational cost of the two binning criteria of
Study A. (a) Accuracy $A_j$ of Eq. (\ref{eq:accuracy}) for the four
mechanical properties, where the markers denote the mean, the bars the
minimum to maximum over the eight bins, and the dashed line the 80\%
level. (b) Mean computational cost per binning criterion denoted by a circle, while the
whiskers span the minimum and maximum Bin costs of each criterion.}
\label{fig:studyA_coverage_cost}
\end{figure}

Fig.~\ref{fig:studyA_coverage_cost} compares the two binning criteria in terms
of accuracy, defined in Eq.~\eqref{eq:accuracy}, and computational cost. In Fig.~\ref{fig:studyA_coverage_cost}a, the markers denote
the mean accuracy over the eight Cells, the bars span the minimum to maximum accuracy value, and the dashed line marks the 80\% reference accuracy level. The mean accuracies exceed 80\% for all four properties under both criteria, but the minimum values differ significantly. With
$S_u$ binning, the minimum accuracies of $S_u$, $E_{50}$, and $p_o$ remain at
or above 80\%, and only $C_c$ falls below it. With $e$ binning, except $p_o$, minimum accuracies of all properties fall below 80\%. Compared to $S_u$ binning, the mean accuracies of $p_o$ and $C_c$ increase slightly, while those of $S_u$ and
$E_{50}$ decrease. Fig.~\ref{fig:studyA_coverage_cost}b reports the computational cost in terms of total ABAQUS runs. Binning by $S_u$ requires 780 runs per Bin on average against 1,071 for binning by $e$, an increase of about 37\%. These results support two conclusions. First, $S_u$ is the more computationally efficient one-level binning criterion. Second, binning by $S_u$ yields higher mean accuracies for $S_u$ and $E_{50}$, while binning by $e$ favors $p_o$ and $C_c$. In Study B, we examine whether the accuracy achieved by binning with $S_u$
alone can be improved by a second binning level based on $e$.

\subsection{Study B: Sub-binning}
\label{sec:studyB}

In Study B, we evaluate the effect of a second binning level. Following the
procedure of Section~\ref{sec:demo}, we partition the working dataset into four
$S_u$ bins and divide each Bin into four $e$ sub-bins, producing 16 Cells. We
select $n_s = 30$ samples per Cell for calibration, giving 480 calibration
dataset samples, of which 471 (98.1\%) converge.

\begin{figure}[H]
  \centering
  \includegraphics[width=\textwidth]{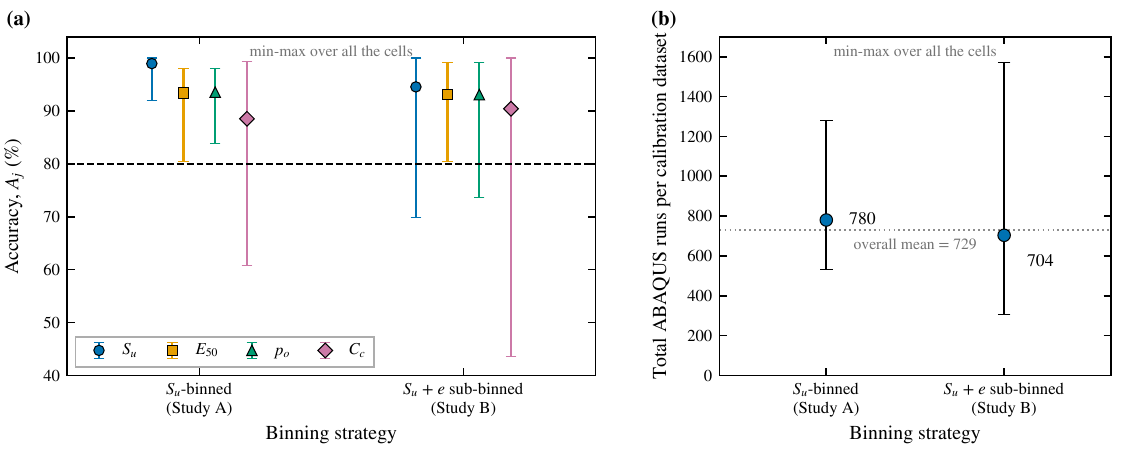}
  \caption{Accuracy and computational cost of the one-level ($S_u$-binned,
  Study A) and two-level ($S_u$-binning + $e$ sub-binned, Study B) binning strategies.
  (a)~Accuracy $A_j$ of Eq.~\eqref{eq:accuracy} for the four mechanical
  properties, where the markers denote the mean, the bars span the minimum to
  maximum over the 8 Cells of Study A and the 16 Cells of Study B, and the
  dashed line the 80\% reference level. (b)~Number of ABAQUS runs per calibration dataset where the mean value is denoted by a circle, while the whiskers span the corresponding
  minimum and maximum values and the dotted line indicates the overall mean of
  729 runs.}
  \label{fig:StudyB_accuracy_cost}
\end{figure}

Fig.~\ref{fig:StudyB_accuracy_cost} compares the one-level and two-level
binning strategies of Study A and B respectively. In Fig.~\ref{fig:StudyB_accuracy_cost}a, the markers
denote the mean accuracy $A_j$ and the bars the minimum to maximum over the
8 Cells of Study A and the 16 Cells of Study B. The mean accuracies of both studies are
comparable where $E_{50}$ and $p_o$ in both cases are within half a percentage point, while
$C_c$ improves from 88.5\% to 90.4\%. The mean accuracy of $S_u$ decreases from 98.9\% (Study A) to 94.5\% (Study B),
which can be explained by the coarser resolution of four $S_u$ bins in Study B instead of eight in Study A. The
wider mean accuracy range of Study B stems from two Cells B4S3 and B4S4 alone while the remaining 14 Cells attain a minimum of $\bar A = 80\%$ for
every property.

Fig.~\ref{fig:StudyB_accuracy_cost}b reports the computational cost in terms of number of ABAQUS runs per
calibration dataset for the two studies, where the circles denote the mean
and the whiskers the minimum to maximum values. The mean cost of Study B is about $10\%$ lower than that of Study A. However, its cost range is wider, from 307 to 1,570 runs compared to 533 to 1,280 runs for Study A. This wider range is largely due to the nine non-converged samples of Study B,
seven of which are concentrated in the Cells of $S_u$ Bin~1 (three in B1S4,
two in B1S1, and one each in B1S2 and B1S3). Excluding the six Cells that
contain non-converged samples, the maximum cost of the remaining ten Cells is
920 runs, and the cost range of Study B becomes 307 to 920 runs. Overall, the two-level binning strategy of Study B divides the working dataset
into 16 Cells, compared to the 8 of the one-level strategy of Study A,
so each calibrated Cell represents a narrower range of material behavior. It achieves this at comparable accuracy and an
approximately 10\% lower mean computational cost per calibration dataset, and
is therefore the preferred binning approach.

\subsection{Study C: Sampling Method and Sample Budget}
\label{sec:studyC}
Here, we examine the sampling methods and sampling budget $n_{s}$ used in the selection of the calibration dataset CS. The study uses Cell B1S4 of Study B, comprising 163 samples. Five sampling methods are compared, namely (a) Puchwein, (b) Latin hypercube sampling (LHS) subset, (c) Sobol subset, (d) Kennard-Stone, and (e) conditioned LHS (cLHS), each at three sample budgets $n_{s}=20$, 30, and 40. All five sampling methods select the calibration samples using only two basic properties, namely the liquid limit $LL$ and water content $w$. No mechanical property is used in the selection process. 
Each of the 15 configurations calibrates its $n_{s}$ calibration set samples and validates against the remaining $163-n_{s}$ validation dataset samples. In total, 450 calibration samples are processed, of which 416 (92.4\%) converge.

\begin{figure}[H]
\centering
\includegraphics[width=0.55\textwidth,trim = 0cm 0cm 0cm 0cm, clip]{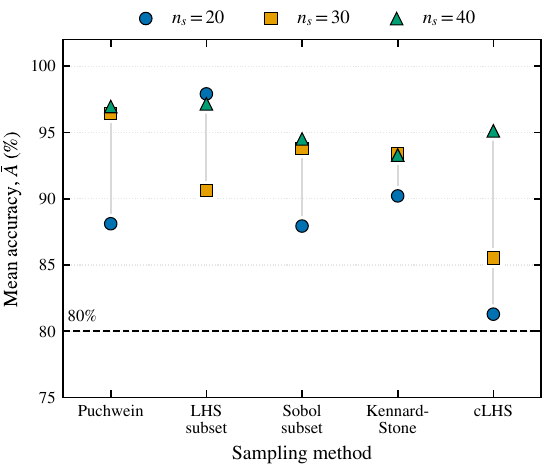}
\caption{Average accuracy of the four mechanical properties for each sampling method at the three sample budgets $n_{s}$. The dashed line marks 80\% coverage.}
\label{fig:studyC_coverage}
\end{figure}

Fig.~\ref{fig:studyC_coverage} shows the average accuracy $\bar A$ of each configuration, i.e., the mean accuracy of the four mechanical properties as defined in Eq.~\ref{eq:accuracy}. Every configuration has an $\bar A$ higher than 81\% but the methods nevertheless differ in their overall performance. A larger budget does not correlate to a better $\bar A$. The LHS subset drops from 97.9\% at $n_{s}=20$ to 90.6\% at $n_{s}=30$, while only in the Puchwein, Sobol subset, and the cLHS cases do the $\bar A$ improve monotonically with the budget. Over the three budgets, the LHS subset and Puchwein attain the highest means, at 95.2 and 93.8\%, and cLHS the lowest at 87.3\%.

\begin{figure}[H]
\centering
\includegraphics[width=1\textwidth,trim = 0cm 0cm 0cm 0cm, clip]{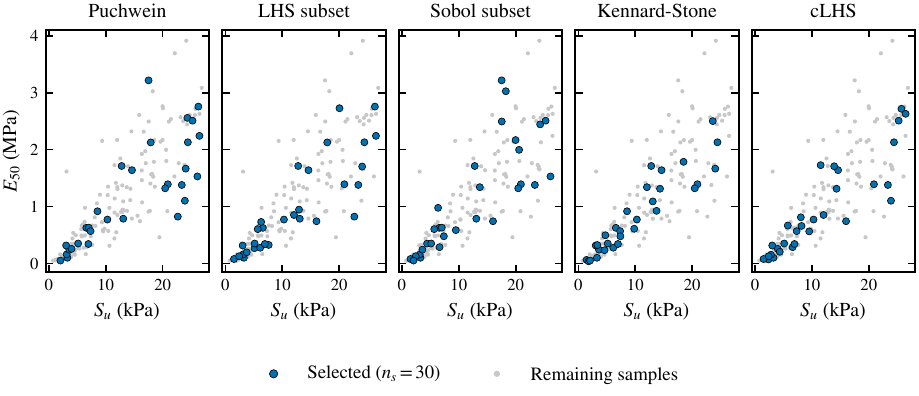}
\caption{Calibration samples selected by each sampling method at $n_{s}=30$, shown in the triaxial target plane. Grey points denote the validation dataset samples of the Cell.}
\label{fig:studyC_selection}
\end{figure}

Fig.~\ref{fig:studyC_selection} helps explain the differences in mean accuracy $\bar{A}$ among the sampling methods. The figure maps the calibration samples selected by each  method at $n_s=30$ onto the triaxial mechanical property space ($S_u$, $E_{50}$). Although the sampling operation uses only the basic properties, this representation allows us  to assess how well the resulting calibration set covers the $S_u$ and $E_{50}$ ranges within the Cell. For brevity, we present only the triaxial plane, although the same  principle applies to the oedometer plane ($p_o$, $C_c$). All five methods span most of the $S_u$ range. However, they differ substantially in the proportion of the $E_{50}$  range covered by the selected samples. The Puchwein and Sobol subset methods each span approximately $82\%$ of the $E_{50}$ range. By contrast, Kennard-Stone spans only $64\%$  and does not select the samples with the highest $E_{50}$ values. Provided that the selected samples calibrate successfully, a broader initial coverage of an mechanical property range generally produces a more representative calibrated CS parameter range and, consequently, a higher corresponding value of $A_j$. This consistent with the trends in  Figs.~\ref{fig:studyC_coverage} and~\ref{fig:studyC_selection}. Based on these results, we use the Puchwein sampling with $n_s=30$ as the reference configuration as it provides a favorable in-Cell property-range accuracy during sampling. The selected configuration represents one suitable option rather than a universally optimal sampling strategy. The BRIDGE framework can accommodate other sampling methods as well based on the coverage requirements and computational budget of the application.

\subsection{Calibration Reliability and Accuracy}
\label{sec:reliability}

The results of the three studies A, B and C give insights into the reliability and the accuracy of 
the BRIDGE framework under different the binning criterion, sub-binning, sampling method, and sample budget. 
Across the 47 evaluated Cells, there are 1,410 calibration instances. These instances 
do not correspond to distinct samples as some calibration datasets overlap across the three studies. 
For example, Study C applies 15 sampling configurations to the same Cell, allowing some samples to 
be selected and calibrated more than once.

\begin{figure}[H]
\centering
\includegraphics[width=1\textwidth,trim=0cm 0cm 0cm 0cm,clip]{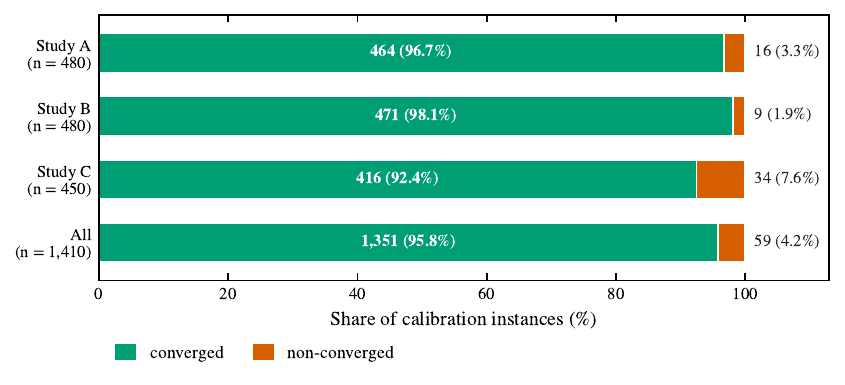}
\caption{Convergence outcomes for the 1,410 calibration instances evaluated in Studies A, B, 
and C. The labels report the numbers and percentages of converged and non-converged instances 
for each study and for the combined results.}
\label{fig:reliability}
\end{figure}

Fig.~\ref{fig:reliability} summarizes the convergence outcomes of the three studies. In Study A $96.7\%$, 464 of the 480, of the  
calibration instances converge. Study B achieves the highest 
proportion, with 471 of 480 instances ($98.1\%$) converging. This result is consistent with 
the improved within-Cell homogeneity provided by the second binning level. Study C has the lowest 
proportion, with 416 of 450 instances ($92.4\%$) converging. Although Study C contains fewer 
calibration instances than either Study A or Study B, it accounts for 34 of the 59 non-converged 
instances. This suggests that certain combinations of sampling method and sample 
budget select samples that are more difficult to calibrate. Overall, 1,351 of the 1,410 
calibration instances converge, giving an overall convergence percentage of $95.8\%$. This overall
convergence percentage represents the high reliability of the calibration procedure across the various evaluated configurations.

\begin{figure}[H]
\centering
\includegraphics[width=1\textwidth,trim=0cm 0cm 0cm 0cm,clip]{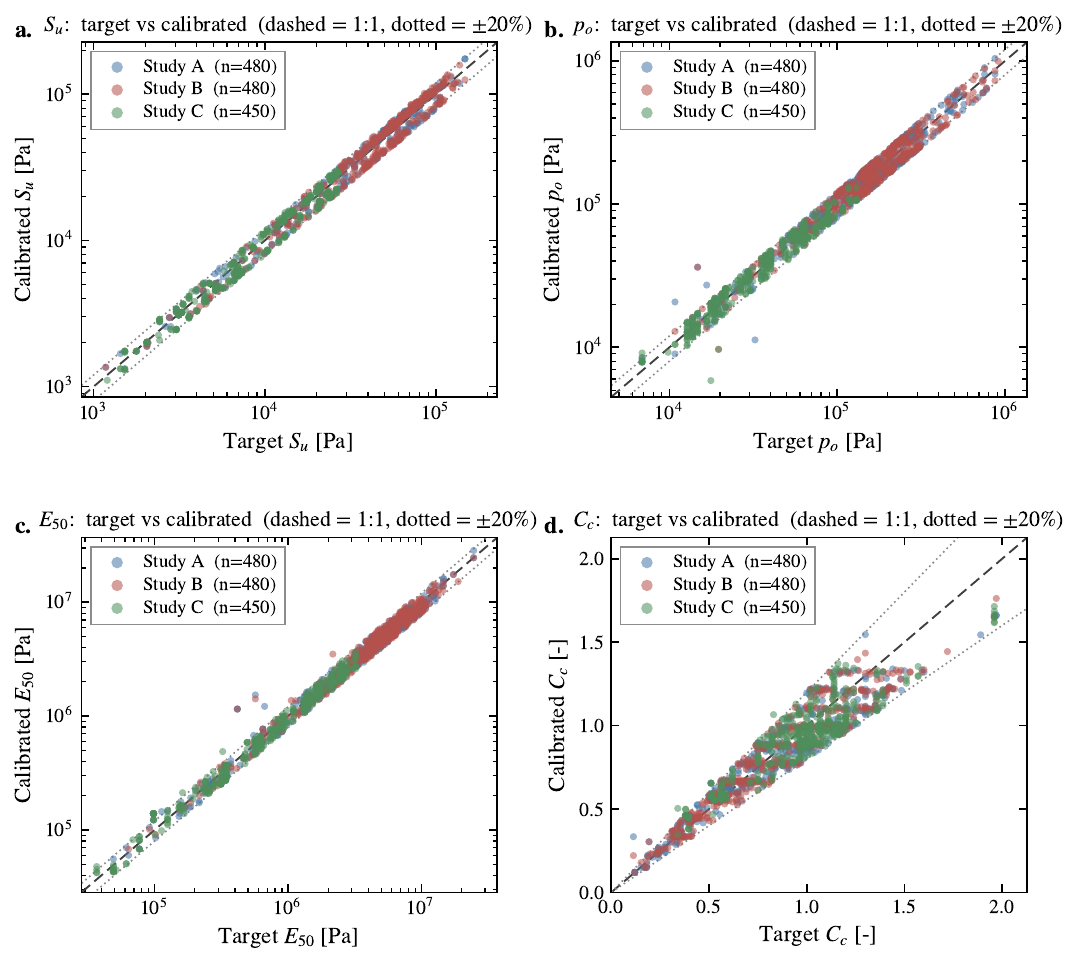}
\caption{Comparison of the calibrated mechanical properties $\mathrm{MS}_{\mathrm{out}}$ with 
their reference values $\mathrm{MS}_{\mathrm{ref}}$ across Studies A, B, and C for (a) $S_u$, 
(b) $p_o$, (c) $E_{50}$, and (d) $C_c$. The dashed lines indicate perfect agreement, and the 
dotted lines define the $\pm20\%$ tolerance band.}
\label{fig:accuracy}
\end{figure}

Fig.~\ref{fig:accuracy} compares the final calibrated mechanical properties 
$\mathrm{MS}_{\mathrm{out}}$ with their reference values $\mathrm{MS}_{\mathrm{ref}}$. 
The dashed line represent perfect agreement, while the dotted lines define the prescribed 
$\pm20\%$ tolerance band. Most calibrated values lie close to the line of perfect agreement 
for all four mechanical properties. The $S_u$ and $E_{50}$ results follow this line closely 
over several orders of magnitude. The $p_o$ and $C_c$ results exhibit greater scatter at the 
lower values, although most remain within the tolerance band. 

\begin{figure}[H]
\centering
\includegraphics[width=1\textwidth,trim=0cm 0cm 0cm 0cm,clip]{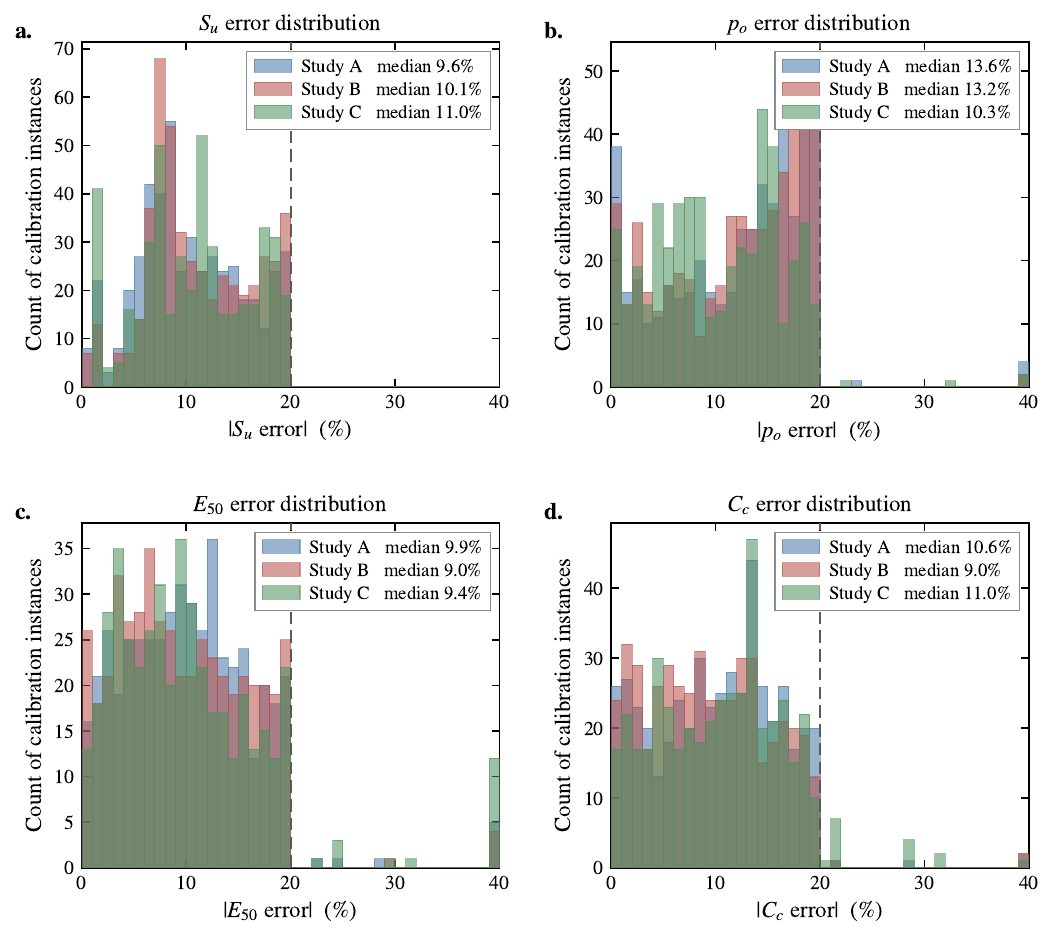}
\caption{Distributions of the absolute calibration errors across Studies A, B, and C for (a) 
$S_u$, (b) $p_o$, (c) $E_{50}$, and (d) $C_c$. The legends report the median error for each 
study, and the dashed lines indicate the prescribed $20\%$ stage tolerance.}
\label{fig:errors}
\end{figure}

Fig.~\ref{fig:errors} presents the distributions of the absolute percentage calibration errors 
for the four mechanical properties across Studies A, B, and C. For each calibration instance, the absolute percentage error of each mechanical property ($S_u$, $E_{50}$, $p_o$, and $C_c$) is calculated as the absolute difference between its calibrated and reference values, divided by the reference value and multiplied by 100. The median errors for $S_u$ range from $9.6\%$ to $11.0\%$, 
while those for $E_{50}$ range from $9.0\%$ to $9.9\%$. The median errors for $p_o$ and $C_c$ range from $10.3\%$ to 
$13.6\%$ and from $9.0\%$ to $11.0\%$, respectively. All study-specific medians remain below 
the prescribed $20\%$ stage tolerance. The similarity of these medians indicates that the typical 
calibration accuracy remains consistent across the three studies.

The error distributions provide additional information that is not captured by 
the medians. The $S_u$ errors are always below the tolerance across all three studies. The 
$p_o$ distributions have higher median errors but contain relatively few large-error instances. 
By contrast, the $E_{50}$ and $C_c$ error distributions have several instances with errors approaching $40\%$. 
Thus, challenges in calibrating these two properties appear to account for most of the non-converged cases across the three studies.
Taken together, the results indicate that the performance of the multi-stage calibration is consistent across the 
binning and sampling configurations considered.

%%%%%%%%%%%%%%%%%%%%%%%%%%%%%%%%%%%%%%%%%%%%%%%%%%%
\section{Conclusions}
\label{sec:conclusions}
 
We presented the BRIDGE framework, a physics-informed tool that addresses the characterization gap between basic and mechanical properties. The framework groups a marine clay database into Cells based on physically meaningful criteria, extracts a small calibration dataset (less than 1/5th the Cell size) from each Cell, and calibrates the Extended Cam-Clay parameters of every calibration sample through a multi stage optimization of single-element finite element simulations. Predictive envelopes are then constructed from the measured mechanical properties of the converged samples, against which the validation samples of the Cell are classified as pass, marginal, or fail. The final output of the framework is the range of the calibrated CS parameters of every Cell, which practitioners can adopt as useful bounds to simulate the response of soils whose basic properties fall within that Cell. Three numerical studies and a reliability assessment demonstrated features of the framework, the trade-offs among them, and the key conclusions of this work are as follows.
\begin{enumerate}
\item \textbf{Binning criterion:} binning by $S_{u}$ matches the coverage of binning by $e$ at a lower computational cost (Study A).
\item \textbf{Second binning level:} sub-binning helps maintain accuracy at lower mean computational cost (Study B).
\item \textbf{Sampling method and sample budget:} mean accuracy is affected by the choice of sampling method and sample budget. Puchwein sampling with $n_{s}=30$ was the configuration adopted in this work, however, this configuration is not presented as universally optimal, and its selection ultimately remains a user-defined choice (Study C).
\item \textbf{Reliability and accuracy:} the calibration converges for 95.8\% of the 1,410 calibration instances with median errors of 9 to 14\%.
(Section~\ref{sec:reliability}).
\item \textbf{Data efficiency:} BRIDGE uses only 10 to 20\% of the total samples to produce a validated mechanical surrogate for the Cell.
\item \textbf{Useful output:} once validated, the mechanical property ranges of a Cell can be used to close the characterization gap between basic and mechanical properties for a new sample assigned to the Cell.
\end{enumerate}

The numerical studies identify several avenues for future research. First, the sampling methods and sample budget can be evaluated across all Cells to determine 
how their performance is influenced by one-level and two-level binning. Second, the concentration of non-converged calibrations in low-$S_u$, low-$e$ Cells requires further investigation. The greater contribution of $E_{50}$ and $C_c$ to non-convergence should also be examined through further analyses. Finally, BRIDGE could also be applied to new soil databases to establish the broader implementation of the framework.

Existing approaches address individual aspects of the characterization gap,
but none simultaneously predicts mechanical properties from routine basic 
properties, maintains physical consistency, and requires only limited advanced
experimental data. BRIDGE closes this gap. The calibration dataset comprises
less than 20\% of the data in every modeling scenario. The physics of the
mechanical model, combined with the bespoke binning and sampling strategies,
allows the framework to learn from a calibration dataset of this limited size.
The resulting predictive envelopes, built from physically consistent finite
element simulations, represent more than 90\% of the validation dataset
samples. Once the mechanical surrogate of a Cell is validated, its mechanical property ranges can be used to predict the $E_{50}$, $p_o$, and $C_c$ ranges for any new sample whose
basic properties and $S_u$ assign it to that Cell. Since every component of the framework is interchangeable, the same architecture extends naturally to other constitutive models, geomaterial databases, and engineering applications.

%%%%%%%%%%%%%%%%%%%%%%%%%%%%%%%%%%%%%%%%%%%%%%%%%%%
% Appendix
\appendix

\section{Constitutive Model: Extended Cam-Clay}
\label{apx:ecc}
 
This section presents the Extended Cam Clay (ECC) constitutive model used in the BRIDGE framework. We also present the four constitutive parameters $\kappa$, $\lambda$, $M$ and $a_0$ which constitute the calibration set (CS) on which the multi-stage calibration operates. We adopt the Extended Cam-Clay model, which generalizes the Modified Cam-Clay formulation of \citep{roscoe1963yielding,roscoe1968generalised}, because of  (a) it is a critical-state soil model  \citep{schofield1968critical}, which is well suited to the marine clays that populate the dataset under consideration~\ref{sec:dataset}, and (b) its availability as the Clay Plasticity model in ABAQUS  \citep{abaqus2024analysis}, which is used in the finite element analysis part of the calibration workflow. In ECC, isotropic compression is idealized as bilinear in the plane of the void ratio $e$ against the logarithm of the mean effective stress $p$,
\begin{equation}
\mathrm{d}e = -\kappa \,\mathrm{d}\!\left(\ln p\right)
\;\; \text{(swelling lines)}
\qquad \text{and} \qquad
\mathrm{d}e = -\lambda \,\mathrm{d}\!\left(\ln p\right)
\;\; \text{(virgin compression line)},
\label{eq:ecc_compression}
\end{equation}
where in ABAQUS, $\kappa$ is called the logarithmic bulk modulus of the porous elasticity definition, while $\lambda$ is called the logarithmic plastic bulk modulus of the Clay Plasticity definition. The critical state, at which shear deformation continues at constant volume and constant effective stress, is characterized by
\begin{equation}
q = M\,p,
\label{eq:ecc_csl}
\end{equation}
where $q$ is the deviatoric stress and $M$ is the critical-state stress ratio. Yielding initiates on the elliptical surface
\begin{equation}
q^{2} = M^{2}\, p \left(2\,a_{0} - p\right),
\label{eq:ecc_yield}
\end{equation}
where $a_{0}$ is the initial yield surface size, which bounds the elastic domain, while plastic flow is associated and the surface expands or contracts with plastic volumetric strain. Further information on the model and its implementation is available in the ABAQUS documentation  \citep{abaqus2024analysis}.

\section{Pearson and Spearman correlations}
\label{apx:correlation_equations}

Here, we present the definitions and expressions for the Pearson and Spearman correlations. The Pearson coefficient quantifies the strength of the linear association between two parameters \citep{pearson1895} and is defined as
\begin{equation}
r=\dfrac{\sum_{i=1}^{n}\left(x_{i}-\bar{x}\right)\left(y_{i}-\bar{y}\right)}
{\sqrt{\sum_{i=1}^{n}\left(x_{i}-\bar{x}\right)^{2}}
\sqrt{\sum_{i=1}^{n}\left(y_{i}-\bar{y}\right)^{2}}},
\label{eq:pearson}
\end{equation}
where $x_{i}$ and $y_{i}$ denote the values of the two parameters recorded for sample $i$, $n$ denotes the number of samples, and $\bar{x}$ and $\bar{y}$ denote the corresponding means over all samples. The The Spearman coefficient is the Pearson coefficient evaluated on the ranks of the values \citep{spearman1904}. The rank $R(x_{i})$ of a value is its position in the ascending order of the $n$ values of the parameter, from 1 for the smallest to $n$ for the largest, with tied values assigned their average rank. Since the ranks retain only the ordering, the Spearman coefficient quantifies the monotonic rather than the linear association and is defined as
\begin{equation}
\rho=\frac{\sum_{i=1}^{n}\left(R(x_{i})-\bar{R}\right)
\left(R(y_{i})-\bar{R}\right)}
{\sqrt{\sum_{i=1}^{n}\left(R(x_{i})-\bar{R}\right)^{2}}
\sqrt{\sum_{i=1}^{n}\left(R(y_{i})-\bar{R}\right)^{2}}},
\label{eq:spearman}
\end{equation}
where $\bar{R}=\left(n+1\right)/2$ denotes the mean rank of either
parameter. The Spearman coefficient quantifies the strength of the monotonic association between the two parameters.

\section{Physical and Numerical Safeguards}
\label{apx:guards}

The calibration procedure of Section~\ref{sec:calibration} has the following safeguards implemented to ensure physical meaningfulness of the CS and numerical stability. 

{\renewcommand{\labelenumi}{(\alph{enumi})}%
\begin{enumerate}
\item The elastic slope is kept below the plastic slope through $\kappa < 0.925\,\lambda$. The 7.5\% gap prevents the degeneration of the preconsolidation response as $\kappa$ approaches $\lambda$.
\item The overconsolidation ratio is fixed at 1.5. A ratio of unity leaves no elastic zone and makes $\kappa$ irrelevant to $E_{50}$.
\item The Poisson's ratio is clamped to $\nu \geq (3 - M)/(6 + M) + 0.02$. This wet-side condition prevents the one-dimensional elastic path of the oedometer simulation from crossing the critical state line and triggering spurious softening.
\end{enumerate}}

These guards convert the most frequent cause of numerical instability into either admissible corrections or explicit non-convergence flags, so that every calibrated
parameter (CS) set that is used for the FEM simulations remains physically meaningful.

%%%%%%%%%%%%%%%%%%%%%%%%%%%%%%%%%%%%%%%%%%%%%%%%%%%
\bibliography{bibliography}
%%%%%%%%%%%%%%%%%%%%%%%%%%%%%%%%%%%%%%%%%%%%%%%%%%%

\end{document}